\documentclass[twocolumn]{aastex701}
\usepackage{amsmath}
\usepackage[T1]{fontenc}
\usepackage{lmodern}

\begin{document}

\title{Multi-wavelength Emission Modeling from Accretion Flows around Isolated Black Holes Including Magnetic Flux Transport}

\author[orcid=0009-0002-3930-4960, gname='Takumi', sname='Koshimizu']{Takumi Koshimizu}
\affiliation{Astronomical Institute, Graduate School of Science,Tohoku University}
\email[show]{koshimizu.takumi.r2@dc.tohoku.ac.jp}  

\author[orcid=0000-0003-2579-7266, gname=Shigeo, sname='Kimura']{Shigeo S. Kimura} 
\affiliation{Frontier Research Institute for Interdisciplinary Sciences, Tohiku University}
\email[show]{shigeo@astr.tohoku.ac.jp}



\begin{abstract}
Isolated stellar-mass black holes (IBHs) are expected to be abundant in the Milky Way, yet their electromagnetic signatures remain largely undetected. 
We investigate the detectability of IBHs in molecular clouds using a 1D, multi-wavelength emission model that incorporates magnetic flux transport controlled by the magnetic Prandtl number $P_m$. 
We find that magnetically arrested disks (MADs) form for $P_m\gtrsim 1$, where the magnetic flux threading the black hole is in a saturation value. 
On the other hand, MAD formation is restricted to a limited parameter range for $P_m<1$, 
In our model, outer parts of accretion disks, around 100 gravitational radii, efficiently emit infrared photons detectable by WISE. This feature is not captured by the conventional one-zone model. 
X-ray emission depends strongly on $P_m$; 
For $P_m=1$ where MAD is formed, X-ray emission is dominated by nonthermal radiation, whereas 
inverse Compton emission becomes dominant 
for $P_m=0.5$ where the magnetic field is weaker than the saturation value. X-ray detection is plausible if they are in dense molecular-cloud filaments within $1\mathrm{kpc}$ for $P_m\ge1$, although it is challenging for $P_m< 1$. 
These results demonstrate that magnetic flux transport plays a key role in shaping the multiwavelength observational signatures of IBHs.
\end{abstract}



\section{Introduction} 

The existence of stellar-mass black holes (BHs) has been confirmed through decades of observations of Galactic X-ray binaries (see \citealt{2016A&A...587A..61C,2016yCat..22220015T}),
increasing numbers of gravitational-wave detections
from compact binary mergers (e.g., \citealt{2023PhRvX..13d1039A}),
and the analysis of the dynamical motion of binaries in Gaia data (Gaia BH 1, 2, and 3; \citealt{2023MNRAS.518.1057E,2023MNRAS.521.4323E,2025OJAp....8E..79T,2024A&A...686L...2G})
Stellar-mass BHs are believed to form as
an end product of stars of initial masses higher than
$\sim 25M_{\odot}$ \citep{2002RvMP...74.1015W}. 
Considering the star formation rate and
age of the universe, there should be roughly $10^8$ BHs in our Galaxy (e.g., \citealt{1983bhwd.book.....S, 1994ApJ...423..659B, 2020A&A...638A..94O}), and the majority of them are expected to be isolated stellarmass BHs (IBHs).
However, identifying IBHs through their electromagnetic emission remains extremely challenging. 
To date, only one firmly confirmed IBH has been reported through microlensing observations 
(OGLE-2011-BLG-0462/MOA-2011-BLG-191; \citealt{2023ApJ...955..116L,2022ApJ...933...83S}). 
Most isolated black holes in the Galaxy are still undetected.

IBHs in the Galactic plane are expected to accrete interstellar
medium (ISM) via the Bondi-Hoyle–Lyttleton (BHL) accretion \citep[see][for a review]{2004NewAR..48..843E}.
The nonzero net angular momentum of infalling gas leads to the formation of an accretion disk around the IBH \citep{2002MNRAS.334..553A, 2017MNRAS.470.3332I}.
In a typical ISM environment, the BHL
accretion rate onto IBHs is so low that their accretion disk is expected to be a radiatively inefficient accretion flow (RIAF; \citealt{1994ApJ...428L..13N,2014ARA&A..52..529Y}).
RIAF achieves high temperature, geometrically thick structure (the scale height $H\sim R$, where $R$ is the cylindrical radius in the accretion flow), and high effective viscosity, leading to a high radial velocity $V_R\sim \alpha (H/R)^2 V_K$, where $V_K$ is the Keplerian velocity and $\alpha$ is the viscous parameter \citep{1973A&A....24..337S}.
Since multiwavelength radiation is emitted from the accretion flow \citep{1997ApJ...489..791M, 2003ApJ...598..301Y, 2021ApJ...923..260P, 2021ApJ...915...31K}, 
it may be possible to identify IBHs through observations of this emission \citep{1998ApJ...495L..85F, 1999ApJ...523L...7A, 2002MNRAS.334..553A, 2003ApJ...596..437C,2012MNRAS.427..589B, 
2018MNRAS.477..791T,
2021ApJ...922L..15K,
2025ApJ...988L..12M,
2025A&A...700A..49M,
2025ApJ...985..251K}.
Multi-wavelength emission from OGLE-2011-BLG-046 is also predicted by \citet{2025ApJ...986..135K}, but observations by Chandra indicate no strong X-ray signals from it \citep{2022ApJ...934...62M,2025MNRAS.541.3709M}.

The radiative properties are strongly dependent on the magnetic field strength in the accretion flow.
Depending on the magnetic flux threading the horizon,
the accretion flows are broadly classified into two accretion states; One is the magnetically
arrested disk (MAD; \citealt{2003PASJ...55L..69N, 2012MNRAS.426.3241N}), in which the magnetic flux threading the central BH is at a saturated level. The other is standard and normal evolution (SANE) disks, in which the large-scale magnetic field does not strongly affect the accretion dynamics.
The magnetic field strength and reconnection processes differ significantly between the MAD and SANE states \citep{2012MNRAS.426.3241N,2020ApJ...900..100R}. 
In particular, MADs can efficiently produce nonthermal electrons by magnetic reconnection \citep{2001ApJ...562L..63Z,2020PhPl...27h0501G,2025ARA&A..63..127S}.
These electrons emit X-rays and MeV gamma rays via synchrotron radiation \citep{2016ApJ...826...77B,2020ApJ...905..178K}. 
As a result, the radiative properties differ substantially between MAD and SANE flows. 
Therefore, it is essential to solve the global transport of magnetic flux in accretion disks in order to determine whether a given system reaches the MAD state.

In this paper, We construct 1D multiwavelength emission model including magnetic flux transport. 
We focus on IBHs embedded in molecular clouds within the Milky Way. In such high-density environments, the accretion rate is enhanced and the resulting electromagnetic emission becomes stronger, improving the prospects for their detection.
Gas in the cloud contains large-scale magnetic fields. These magnetized gas is captured by the IBH and forms an accretion disk with the large-scale field.
Depending on the efficiency of magnetic flux transport, the disk evolves into either the SANE or MAD state.
The transported magnetic flux determines the magnetic field strength in the accretion flow, which in turn affects electron heating and the resulting radiation.
Thermal electrons are heated to relativistic
temperatures by dissipation of magnetic energy, 
and they emit infrared and optical signals through thermal synchrotron radiation.
They also produce X-rays through inverse Compton scattering (IC).
In the MAD state, radiation from nonthermal electrons contributes to the multiwavelength photon spectra significantly. 
We evaluate the detectability of these multiwavelength signals with WISE  \citep{2010AJ....140.1868W}, Gaia \citep{2021A&A...649A...1G}, and eROSITA \citep{2021A&A...647A...1P}.
This enables a systematic assessment of IBH detectability across multiple wavelength bands.

This paper is organized as follows. 
In Section~\ref{sec:model}, we describe the accretion flow model for IBHs. 
In Section~\ref{sec:mad_criterion}, we calculate the magnetic flux transport in the accretion disk and evaluate the condition for MAD formation. 
In Section~\ref{sec:radiation}, we compute the multiwavelength spectra from accretion flows in molecular clouds.
In Section~\ref{sec:discussion}, we discuss the implications and caveats of our results. Finally, we summarize our results in Section~\ref{sec:summary}. 
Throughout this paper, we adopt the notation $Q_x \equiv Q/10^x$ in cgs units, except for the BH mass, which is expressed as $M = 10^x M_x M_{\odot}$.

\section{Accretion Flows onto IBHs in Molecular Clouds} \label{sec:model}

We consider an IBH that accretes ambient gas. The effective mass accretion rate can be reduced relative to the classical BHL accretion rate due to bow shock formation, effects of magnetic fields, and kinetic feedback by outflows.
We introduce a parameter, $\lambda_{\mathrm{m}}$,
to take into account the reduction of mass accretion rate.
We here use $\lambda_{\mathrm{m}}=0.10$ as a reference value based on hydrodynamic simulations \citep{2014ApJ...783...50L,2022A&A...660A...5B,2025PhRvD.111h3025K,2025A&A...694A.239J}.
Then, we estimate the mass accretion rate at the outer edge of the accretion disk as
\begin{equation}
\begin{split} 
\dot{M}_0 & \approx
\lambda_{\mathrm{m}}\frac{4\pi G^2M^2\mu_{\mathrm{MC}}m_p n_{\mathrm{MC}}}{V_{\mathrm{eff}}^3} \\
& \simeq 1.1\times 10^{15}\lambda_{m,-1} M_1^2n_{\mathrm{MC},3}V_{\mathrm{eff},6.3}^{-3}\;\;\mathrm{g\,s^{-1}},
\end{split}
\label{eq:Mdot_0}
\end{equation}
where $G$ is the gravitational constant, $M$ is the IBH mass,
$m_p$ is the proton mass, $V_{\mathrm{eff}}$ is the relative velocity between IBH and the molecular gas.
Here, $\mu_{\mathrm{MC}}=2.3$ and $n_{\mathrm{MC}}$ are the mean molecular weight and number density of the molecular gas, respectively.

\begin{figure}[t]
\plotone{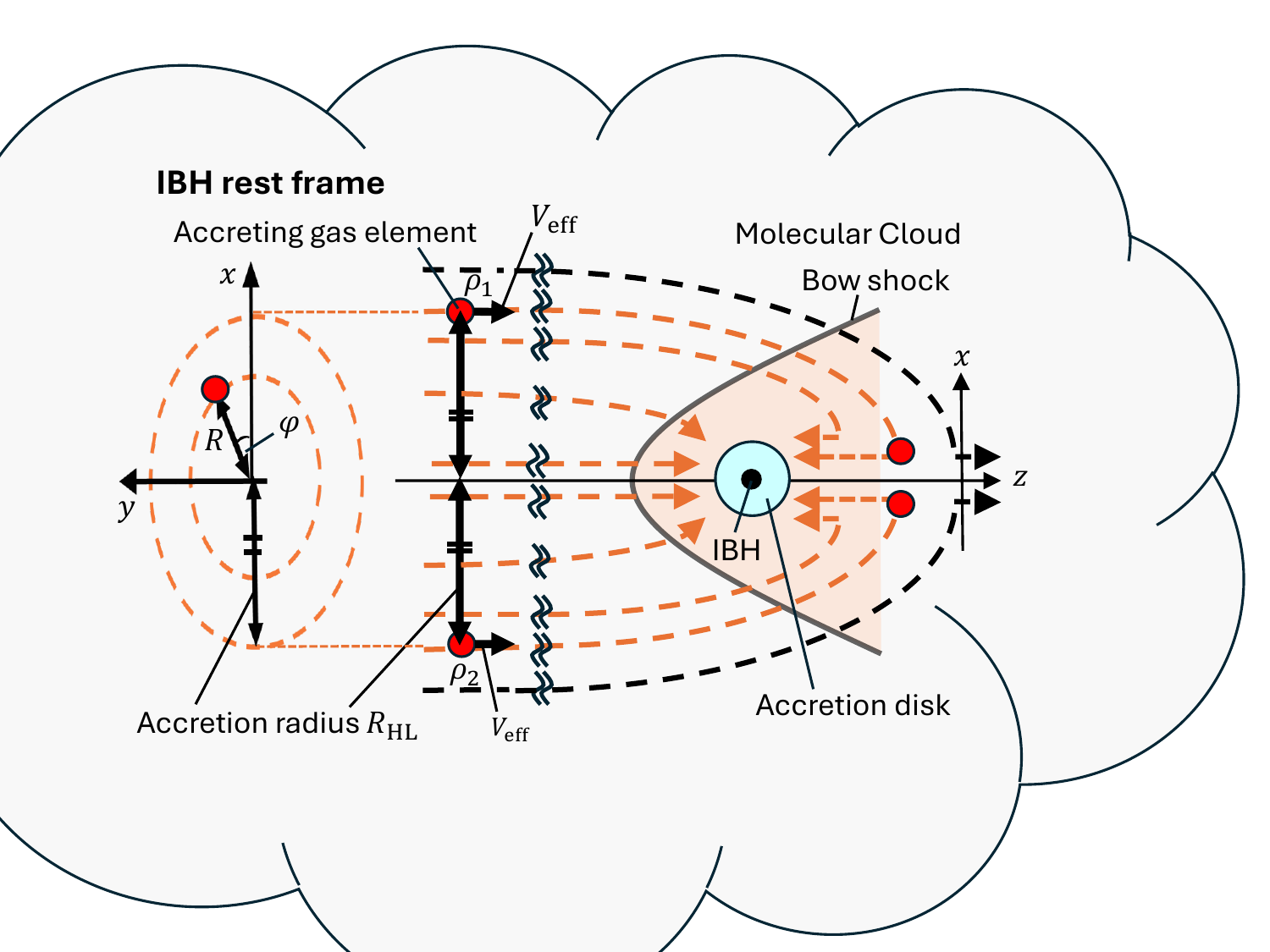}
\caption{Schematic picture of accretion disk formation around an IBH moving through a molecular cloud.
In IBH rest frame, molecular gas flows in from infinity along the z-axis with the velocity $V_{\mathrm{eff}}$, forming a bow shock owing to the supersonic motion of the IBH. 
The dashed lines represent stream lines of the flow.
The orange and black lines indicate captured and unbound gas trajectories, respectively.
If the molecular cloud has a density gradient in the $x$-direction, the captured gas acquires net angular momentum and forms an accretion disk whose angular momentum is aligned with the $y$-direction.
\label{fig:accretion_disk}}
\end{figure}

Figure \ref{fig:accretion_disk} indicates a schematic picture of the formation of an accretion disk.
When the density gradient is aligned to the x-direction,
the accreting matter acquires a net specific angular momentum $\boldsymbol{j}\sim 0.25\,\delta\rho/\rho|_{L=R_{\mathrm{BHL}}}V_{\mathrm{eff}}R_{\mathrm{BHL}}\boldsymbol{e}_y$,
where $\rho$ is the average density, and 
$\boldsymbol{e}_y$ is the unit vector in the y-direction 
\citep{2002MNRAS.334..553A, 2017MNRAS.470.3332I}.
The molecular cloud density has a turbulent fluctuation with a power-law spectrum $\delta\rho/\rho\sim\left[ L/(6\times 10^{18}\mathrm{cm})\right]^{1/3}$
\citep{1995ApJ...443..209A,2011piim.book.....D}.
The accreted matter is circularized before falling to the IBH.
By equating $|\boldsymbol{j}|$ with the Keplerian angular momentum $l_{\mathrm{K}}=\sqrt{GMR_{\mathrm{out}}}$,
we obtain the radius of the resulting accretion disk,
\begin{equation}
\frac{R_{\mathrm{out}}}{R_{G}}=1.3\times 10^5 M_1^{2/3} V_{\mathrm{eff},6.3}^{-10/3},
\label{eq:r_out}
\end{equation}
where $R_G=GM/c^2$ is the gravitational radius.

The circularized gas forms an accretion disks and eventually accretes onto the IBH. 
We estimate the mass accretion rate normalized by the Eddington rate,
$\dot{m} 
=\dot{M}c^2/L_{\mathrm{Edd}}
\le \dot{M}_0c^2/L_{\mathrm{Edd}}
\approx 7.6\times 10^{-4} \lambda_{m,-1} M_1 n_{\mathrm{MC},3}V_{\mathrm{eff},6.3}^{-3}$.
With such a low Eddington ratio, 
we expect the accretion flow to be in the RIAF regime
\citep{2014ARA&A..52..529Y}.
MHD simulations revealed that RIAFs ubiquitously produce outflows \citep{2012ApJ...761..129Y,Sadowski2013MAD}.
Consequently, the mass accretion rate in the accretion disk depends on the disk radius, $R$, and can be written as 
\begin{equation}
\dot{M}=\left(\frac{R}{R_{\mathrm{out}}}\right)^{s_w}\dot{M}_0,
\label{eq:Mdot}
\end{equation}
where $s_w$ is a parameter that describes the outflow efficiency
\citep{1999MNRAS.303L...1B}.
MHD simulations suggest $0.4\lesssim s_w\lesssim 0.8$ \citep{2012ApJ...761..129Y,Sadowski2013MAD,Manikantan2024,2025PhRvD.112l3044L}, and we adopt $s_w=0.5$ as a fiducial value.

In RIAFs, the matter cannot cool efficiently within the infall timescale, resulting in a proton temperature comparable to the virial temperature. RIAFs are geometrically thick because of the strong thermal pressure. The
thick geometry allows the development of large turbulent eddies, leading
 to enhanced turbulent viscosity. 
 As a result, angular momentum transport is efficient, yielding a radial inflow velocity faster than that in standard thin disk.
The radial velocity, sound velocity, proton temperature, and number density in RIAFs can be analytically estimated to be 
(see \citealt{2019MNRAS.485..163K, 2020ApJ...905..178K}, for values for active galactic nuclei (AGN))
\begin{align}
V_R &\approx \frac{1}{2}\alpha V_K=8.4\times10^8 r_{0.5}^{-1/2}\alpha_{-1}\;\;\mathrm{cm\,s^{-1}}, \label{eq:RIAF_V_R}\\
C_S &\approx \frac{1}{2}V_K=8.4\times10^{9}r_{0.5}^{-1/2}\;\;\mathrm{cm\,s^{-1}}, \label{eq:RIAF_C_S}\\
k_BT_p &\approx \frac{GMm_p}{4R}=7.3\times10^7r_{0.5}^{-1}\;\;\mathrm{eV}, 
\label{eq:RIAF_T_p}\\
N_p &\approx \frac{\dot{M}}{4\pi RHV_R m_p} \notag\\
&=5.6\times10^{15}\times(4.1\times10^4)^{-s_w}r_{0.5}^{s_w-3/2}\lambda_{m,-1} \notag\\
&\times\alpha_{-1}^{-1}M_1^{-2s_w/3}V_{\mathrm{eff},6.3}^{(10s_w-9)/3}n_{\mathrm{MC},3}\;\;\mathrm{cm^{-3}},
\label{eq:RIAF_N_p}
\end{align}
where $V_K=\sqrt{GM/R}$ is the Keplerian velocity, $\alpha$
is the viscous parameter, and $r=R/R_G$.
Within this formalism, the scale height is estimated to be $H\approx R/2$, consistent with the geometrically thick structure of RIAFs ($H\sim R$).

\section{Magnetic Flux Transport } \label{sec:mad_criterion}

RIAFs can be classified into either a SANE or MAD state, depending on the efficiency of magnetic flux transport
We first estimate the magnetic flux supplied from the molecular cloud to the outer edge of the accretion disk,
and then solve the global magnetic flux transport in the disk to determine whether the flow reaches the MAD state. 

\subsection{Magnetic Flux Transport from Molecular Clouds to Accretion Disks}

The IBH moves supersonically in the molecular cloud, forming a bow shock around the IBH.
In the shocked and unshocked regions, the temperatures and ionization states differ, and therefore the dominant magnetic flux transport mechanisms also differ.
The magnetic field in the unshocked region, which is weakly ionized plasma, is dominated by ambipolar diffusion 
\citep{2011piim.book.....D}.
However, since the gas is not significantly compressed before reaching the bow shock, 
the magnetic field remains close to the ambient molecular cloud value. 
In the shocked region, 
the gas temperature becomes $\gtrsim10^4$ K, and the gas is collisionally ionized, enabling us to use the flux-freezing approximation. 
To further justify this assumption, we examine whether radiative loss can affect the ionization state of the gas in the shocked region.
The radiative loss timescale is given by using a representative value of the cooling function at $T=2\times10^4$--$10^5\,\mathrm{K}$,
$\Lambda/n_e^2 \sim 10^{-22}\,\mathrm{erg\,cm^{3}\,s^{-1}}$
\citep{2011piim.book.....D},
where $n_e$ is the electron number density and $\Lambda$ is the radiative cooling rate per unit volume.
Assuming a strong shock $n_e=4n_{\mathrm{MC}}$, the radiative loss timescale is estimated to be
$t_{\mathrm{loss}} =u/\Lambda\sim
4.2\times10^7\,n_{\mathrm{MC},3}^{-1}T_{4.3}
\,\mathrm{s}$,
where $u$ and $T$ are the internal energy density and temperature in the shocked region, respectively. 
The accretion timescale is
$t_{\mathrm{acc}}
=R_{\mathrm{BHL}}/V_{\mathrm{eff}}=3.4\times10^8\,M_1V_{\mathrm{eff},6.3}^{-3}
\,\mathrm{s}$,
which is longer than the radiative loss timescale, suggesting that radiative loss may become important in the shocked region.
However, the ionization radius $R_{\mathrm{IF}}$ exceeds the accretion radius $R_{\mathrm{BHL}}$ (see Equation \ref{eq:R_IF} and Table~\ref{tab:model}).
Therefore, the accreting gas is expected to be photoionized by radiation from the accretion flow before reaching the black hole, supporting the validity of the flux-freezing approximation adopted in this work.
Consequently, the magnetic field strength scales as $R^{-2}$ until it form rotationally supported accretion disk. 
We estimate the magnetic field strength at the outer edge of the accretion disk to
\begin{equation}
\begin{split}
B_{\mathrm{out}}
&\approx \lambda_0 B_{\mathrm{MC}}\left(\frac{R_{\mathrm{BHL}}}{R_{\mathrm{out}}}\right)^2 \\
&\simeq 23\,\lambda_{0,-0.77}M_1^{-4/3}V_{\mathrm{eff},6.3}^{8/3}n_{\mathrm{MC},3}^{13/20}\;\; \mathrm{G},
\end{split}
\label{eq:B_out}
\end{equation}
where $B_{\mathrm{MC}}=11\,n_{\mathrm{MC},3}^{13/20}\, \mathrm{\mu G}\,(n_{\mathrm{MC}}>200\,\mathrm{cm^{-3}})$ 
\footnote{
For $n_{\mathrm{MC}}<200\,\mathrm{cm^{-3}}$,
the magnetic field strength in molecular clouds is observed to remain approximately constant at
$B_{\mathrm{MC}}\simeq 5\,\mu{\rm G}$
\citep{2010ApJ...725..466C,2011piim.book.....D}.
}
is the magnetic field strength in the molecular cloud
\citep{2010ApJ...725..466C,2011piim.book.....D}. 
Here, $\lambda_0$ is defined as
\begin{equation}
\lambda_0\equiv \frac{\int_0^{R_{\mathrm{HL}}}\left(L_0/R_{\mathrm{HL}}\right)^2 2\pi\zeta d\zeta}{\int_0^{R_{\mathrm{HL}}}2\pi\zeta d\zeta},
\end{equation}
where $\zeta$ is the impact parameter, 
$L_0$ is the distance from the IBH to the bow shock.
We can solve the orbit of the HL accretion flow if we assume a constant density (see Equation (9) of \citealt{2004NewAR..48..843E}). Using the orbit and the shock-front shape
\footnote{In a coordinate system centered on the IBH,
the shock-front shape is approximated as
$x=\sqrt{1.2R_{\mathrm{HL}}(z+0.1R_{\mathrm{HL}})}$ 
based on the hydrodynamic simulation results of
\citet{2025A&A...694A.239J}.}, 
we obtain $\lambda_0=0.17$.

\subsection{Magnetic Flux Transport in Accretion Disks}
\label{sec:sub32}

To model magnetic flux transport in the accretion disk, we adopt the mean-field model for large-scale poloidal magnetic fields developed by \citet{1994MNRAS.267..235L}.
We treat both $\boldsymbol{B}$ and $\boldsymbol{u}$ as azimuthally averaged quantities.
We neglect the conversion of toroidal into poloidal magnetic fields by turbulence (the $\alpha$ dynamo; see, e.g., \citealt{2011ApJ...728..130G}) and consider only the transport of large-scale poloidal magnetic flux.
Under these assumptions, the toroidal magnetic field does not appear in  the induction equation for the poloidal field.

Recent two-dimensional calculations have shown that the vertically averaged model may overestimate the transported large-scale magnetic field by about one order of magnitude in geometrically thick RIAFs with $h=H/R\sim0.5$ \citep{2024MNRAS.530.1218Y}.
However, the magnetic field adopted for the radiation calculation is additionally limited by physical constraints, whose effects are discussed in Section~\ref{sec:subsec41}.

We express the poloidal field in terms of a flux function $\psi$ defined as $\boldsymbol{B}=\nabla\times (\psi\boldsymbol{e}_{\varphi}/R)$. 
The radial and vertical components of the field are written as
\begin{align}
B_R &= -\frac{1}{R}\frac{\partial\psi}{\partial z}, \label{eq:B_R}\\
B_{z} &= \frac{1}{R}\frac{\partial\psi}{\partial R}. \label{eq:B_z}
\end{align}
The flux function is related to the magnetic flux $\Phi$
threading the disk within radius $R$, i.e.,
\begin{equation}
\Phi = 2\pi\int_0^{R}B_z(R',0)R'dR'=2\pi\psi(R).
\end{equation}
The time evolution of the magnetic flux function is governed by the induction equation. 
In terms of $\psi$, this can be written as (see Equation (6) of \citealt{2014ApJ...785..127O})
\begin{equation}
\frac{\partial\psi}{\partial t} = -u_*\frac{\partial\psi}{\partial R} - \frac{2\pi R\eta_*}{cH}K_{\varphi},
\label{eq:psi_time_1d}
\end{equation}
where $1/\eta_*\equiv (1/2H)\int_{-H}^H(1/\eta)dz$,
$u_*\equiv (\eta_*/2H)\int_{-H}^H(u_R/\eta)dz$,
$u_R$ is the radial velocity, $\eta$ is the magnetic diffusivity, 
$K_{\varphi}\equiv \int_{-H}^{H}J_{\varphi}dz$ is the azimuthal component of the surface current, and $J_{\varphi}=(c/4\pi)(\nabla\times\boldsymbol{B})_{\varphi}$
is the azimuthal current density.
Here, we adopt $u_*=V_R$ given by Equation~\eqref{eq:RIAF_V_R}.
To characterize the competition between magnetic advection and diffusion, we introduce the magnetic Prandtl number,
$P_m\equiv\nu/\eta_*$,
where $\nu=\alpha C_S H$ is the kinematic viscosity. For $P_m<1$, magnetic diffusion dominates magnetic-flux transport, whereas for $P_m>1$, advection is dominant.
The flux function is related to $K_{\varphi}$ via Biot–Savart’s law (see Equation (12) and (13) of \citealt{2014ApJ...785..127O}) as
\begin{align}
\psi &= \int Q(R_<,R_>)K_{\varphi}(R')dR', \label{eq:psi_d}\\
Q&=\frac{4}{c}R_>\left[ K\left( \frac{R_<}{R_>}\right)-E\left( \frac{R_<}{R_>}\right)\right],
\end{align}
where $R_< = \mathrm{min}(R,R')$, $R_> = \mathrm{max}(R,R')$,
and $K(x)$ and
$E(x)$ are the complete elliptic integrals defined by $K(x)=\int_0^{\pi/2} (1-x^2\sin^2\theta)^{-1/2}d\theta$ and $E(x)=\int_0^{\pi/2} (1-x^2\sin^2\theta)^{1/2}d\theta$.

The steady-state solution can be analytically estimated to be
 (see Appendix A of \citealt{2014ApJ...785..127O}
 \footnote{Equations \eqref{eq:psi_ana} and \eqref{eq:C_q} are equivalent to Equations (A3) and (A7) of \citet{2014ApJ...785..127O}.
 Note $D\approx \eta_*/V_RH\approx 1/P_m$, using Equations \eqref{eq:RIAF_V_R} and \eqref{eq:RIAF_C_S}.}) 
\begin{align}
\psi &= \psi_{\mathrm{out}}(q)
\left(\frac{R}{R_{\mathrm{out}}}\right)
^{2-q}, \label{eq:psi_ana} \\
\frac{1}{P_m} &= (2-q)C(q),
\label{eq:C_q}
\end{align}
where
\begin{equation}
C(q) \equiv \frac{2}{\pi}\int_0^{2\pi} \left[K(x)-E(x)\right]\left(x^{-q} + x^{3-q}\right)dx.
\end{equation}
In the limiting cases, 
the steady-state solution approaches
$\psi\propto R^2$ for $P_m\ll1$,
corresponding to an approximately constant vertical magnetic field because of the efficient diffusion.
In contrast, for $P_m\gg1$, magnetic flux is nearly frozen into the flow, yielding
$\psi\propto R^0$.
To determine the normalization of the flux function,
$\psi_{\mathrm{out}}$,
we require the magnetic field strength at the outer boundary to satisfy
$B=\sqrt{B_R^2+B_z^2}=B_{\mathrm{out}}$.
Here, we estimate $B_R$ from Ampère's law as
\begin{equation}
B_R = \frac{4\pi}{c}K_{\varphi} + \int_{-H}^H \frac{\partial B_z}{\partial R}dz 
\approx 2P_mB_z+2H\frac{\partial B_z}{\partial R}.
\label{eq:B_R_ver2}
\end{equation}
The first term of Equation~\eqref{eq:B_R_ver2} 
is calculated using Equations~\eqref{eq:B_z} and \eqref{eq:psi_time_1d}. 
In this work, we neglect the $z$-dependence of $\partial B_z/\partial R$.
Using Equations~\eqref{eq:B_z}, \eqref{eq:psi_ana}, and \eqref{eq:B_R_ver2},
we obtain
\begin{equation}
\begin{split}
&\psi_{\mathrm{out}}(q)=f_{\mathrm{out}}(q)B_{\mathrm{out}}R_{\mathrm{out}}^2,\\
&f_{\mathrm{out}}(q)=\frac{1}{(2-q)\sqrt{1+4(P_m-hq)^2}}, 
\end{split}
\end{equation}
where $h= H/R$ is the aspect ratio of the disk.
This power-law solution is applicable to the sonic point, inside which the flow rapidly plunges into the BH and the power-law RIAF solution (Equations~\ref{eq:RIAF_V_R}--~\ref{eq:RIAF_T_p}) is no longer applicable. 
Therefore, we set the innermost radius of the disk to $R_{\mathrm{in}}=3R_G$, following \citet{2011ApJ...737...94C,2012MNRAS.426.3241N}.

The magnetic field strength obtained by the method above could be unrealistically strong. In reality, the magnetic field strength is regulated by the dynamics of the accretion flow, because the accretion is completely halted if the magnetic field threading the disk is too strong. Thus, we need to impose a saturation value of magnetic field, $B_{\rm sat}$, as a function of radius. We consider that the dynamics is regulated by plasma beta, $\beta=8\pi N_p k_BT_p/B^2$. The lower limit can be set by considering the MAD state in which the magnetic field threading the horizon is in a saturation value: $\Phi_{\rm MAD}=50R_G\sqrt{\dot M(R_{\rm in})c}$ \citep{2012MNRAS.426.3241N,2012MNRAS.423.3083M,2020ApJ...900..100R, 2022ApJ...924L..32R}. Since the magnetic flux and mass accretion rate are conserved at $R\le R_{\mathrm{in}}$ \citep[c.f.,][]{2013MNRAS.436.3856S, 2015ApJ...804..101Y,2025arXiv251025842J}, we can estimate the saturation value of plasma beta:
\begin{equation}
\beta_{\mathrm{sat}}=\frac{32\pi^3R_{\rm in}^4 N_p(R_{\mathrm{in}})k_BT_p(R_{\mathrm{in}})}{\Phi_{\rm MAD}^2}\simeq1.7 \alpha_{-1}^{-1}r_{\rm in,0.48}^{3/2},
\label{eq:beta_sat}
\end{equation}
where $r_{\rm in}=R_{\rm in}/R_G$ and we use $\Phi_{\rm MAD}\approx \Phi_H=2\pi R_{\mathrm{in}}^2B_{\mathrm{sat}}$. Assuming that the accretion process is halted if the accretion flow satisfies $\beta < \beta_{\rm sat}$ somewhere in the disk, the upper limit of the magnetic field strength is imposed as 
\begin{equation}
B_{\mathrm{sat}}=\sqrt{\frac{8\pi N_pk_BT_p}{\beta_{\rm sat}}}.\label{eq:Bsat}
\end{equation}

We solve Equations \eqref{eq:psi_time_1d} and \eqref{eq:psi_d} using the same numerical method as \citet{2024MNRAS.530.1218Y} until the steady state is achieved.
In our IBH accretion model, we adopt $\alpha=0.1$, motivated by observations of BH binaries \citep{2019NewA...70....7M}.
We take the computational domain
to be $R_{\mathrm{in}}\le R\le R_{\mathrm{out}}$.
The domain is divided into 512 logarithmically spaced cells.
We use an outflow boundary condition for the inner boundary
\footnote{Regarding the inner boundary conditions, since $u_*$ is always directed inward and the upwind scheme is taken, the choice of inner boundary conditions does not affect the calculation results.}, 
while for the outer boundary we impose $\psi(R=R_{\mathrm{out}},t)
=\psi_{\mathrm{out}}$.
We set the initial condition as
\begin{equation}
\psi(R,t=0) = \psi_{\mathrm{out}}
\left(\frac{R}{R_{\mathrm{out}}}\right)^2,
\end{equation}
which corresponds to a spatially uniform magnetic field, $B=\sqrt{B_R^2+B_z^2}=B_{\mathrm{out}}$.
We confirm that the choice of initial condition does not affect the resulting steady state solution.

\begin{figure}[t]
\plotone{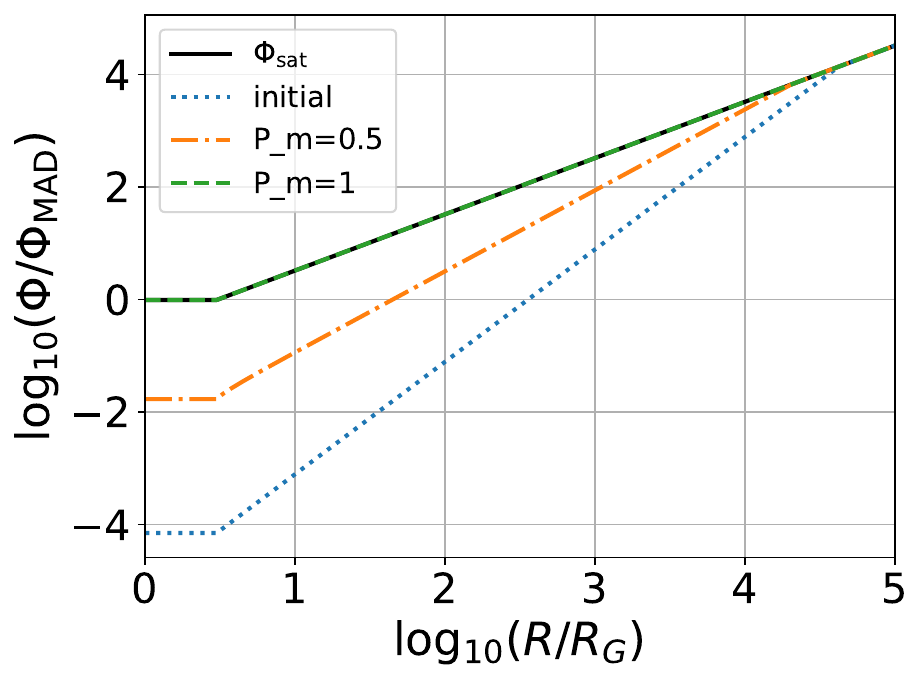}
\caption{The numerical solutions of Equations \eqref{eq:psi_time_1d} and \eqref{eq:psi_d} after the system has reached a steady state 
with $V_{\mathrm{eff}}=20\,\mathrm{km\,s^{-1}}$ 
and $n_{\mathrm{MC}}=10^3\,\mathrm{cm^{-3}}$. 
The solid black and dotted blue lines represent the limiting magnetic flux and the initial condition, respectively. 
The dot-dashed and dashed lines show the steady-state numerical solutions for
$P_m=0.5$ and $P_m=1$,
respectively.
\label{fig:psi_time_result}}
\end{figure}

Figure~\ref{fig:psi_time_result} shows the numerical solutions of Equations~\eqref{eq:psi_time_1d} and \eqref{eq:psi_d} after the system has reached a steady state. 
The magnetic field  at $R=R_{\mathrm{out}}$ is $B_{\mathrm{out}}=23\,\mathrm{G}$ (see Equation \ref{eq:B_out}).
We calculate the cases $P_m=0.5$ and $1$,
since Magnetohydrodynamic (MHD) simulations of MRI-driven turbulence \citep{2009A&A...507...19F} show $0.5\le P_m\le 3$.
MHD simulations and analytic studies also suggest $P_m\sim1$ \citep{2020A&A...636A..93K}.
The magnetic flux function $\psi$ follows a power-law profile $\psi\propto R^s$. 
For $P_m=0.5$, the exponent $s=2-q=1.45>1$, implying that a significant fraction of the magnetic flux escapes before reaching the BH.
In contrast, for $P_m=1$, the exponent decreases to $s=1$, indicating that the diffusive loss of magnetic flux is less efficient and that a larger fraction of the flux is transported toward the BH.
These trends are consistent with the analytical expectations of \citet{2014ApJ...785..127O}.
This numerical solution is further constrained by the limiting magnetic flux $\Phi_{\mathrm{sat}}$, corresponding to $\beta=\beta_{\mathrm{sat}}$.
Figure \ref{fig:psi_time_result} shows that the solution is regulated by $\Phi_{\mathrm{sat}}$ over the entire radial range for $P_m=1$.
This means that the solution should be converged to the same one even if we use a higher value of $P_m\gtrsim1$, and thus, the magnetic-field regulated accretion should be realized in the entire radial range for $P_m\gtrsim1$.

\subsection{MAD criterion}

Using the method described above, we can estimate the magnetic flux at the event horizon of the BH, 
$\Phi_{\mathrm{H}} = 2\pi\psi(R_{\mathrm{H}})$.
The condition for MAD formation is that $\Phi_{\mathrm{H}}$ exceeds the MAD magnetic flux,
$\Phi_{\mathrm{MAD}} = 50 R_G \sqrt{\dot{M}(R_{\mathrm{in}}) c}$.
Using Equation \eqref{eq:psi_ana},
the magnetic flux at the event horizon scales as $\Phi_{\mathrm{H}}\simeq 2\pi \psi_{\mathrm{out}}(R_{\mathrm{in}}/R_{\mathrm{out}})^{2-q}\propto n_{\mathrm{MC}}^{13/20}M^{2(q+1)/3}V_{\mathrm{eff}}^{2(4-5q)/3}r_{\mathrm{in}}^{2-q}$, whereas the MAD magnetic flux scales as 
$\Phi_{\mathrm{MAD}}\propto n_{\mathrm{MC}}^{1/2}M^{(6-s_w)/3}V_{\mathrm{eff}}^{(10s_w-9)/3}r_{\mathrm{in}}^{s_w/2}$.
Thus, the MAD condition $\Phi_{\mathrm{H}}\ge\Phi_{\mathrm{MAD}}$
gives
\begin{equation}
\begin{split}
n_{\mathrm{MC}}\ge 5.3\times 10^{-25}
\times(4.3\times 10^4)^{\frac{10(4-2q-s_w)}
{3}}
f_{\mathrm{out}}(q)^{-\frac{20}{3}} \times
\\
M_1^{\frac{20(4-2q-s_w)}{9}}
V_{\mathrm{eff},6.3}^{\frac{50(-5+4q+2s_w)}
{9}}
r_{\mathrm{in},0.48}^{\frac{10(-4+2q+s_w)}
{3}}.
\label{eq:mad_criterion}
\end{split}
\end{equation}
For $q<1$, the exponent of $V_{\mathrm{eff}}$ in Equation~\eqref{eq:mad_criterion} with $s_w=0.5$ becomes negative, implying that the required density for MAD formation decreases with increasing $V_{\mathrm{eff}}$.
Physically, a larger $V_{\mathrm{eff}}$ reduces the disk size, $R_{\mathrm{out}}\propto V_{\mathrm{eff}}^{-10/3}$, thereby shortening the radial region over which magnetic flux can escape through diffusion.
In contrast, for $q>1$, MAD formation becomes easier at smaller $V_{\mathrm{eff}}$, although MAD is always formed in typical molecular cloud environments.

\begin{figure}[t]
\plotone{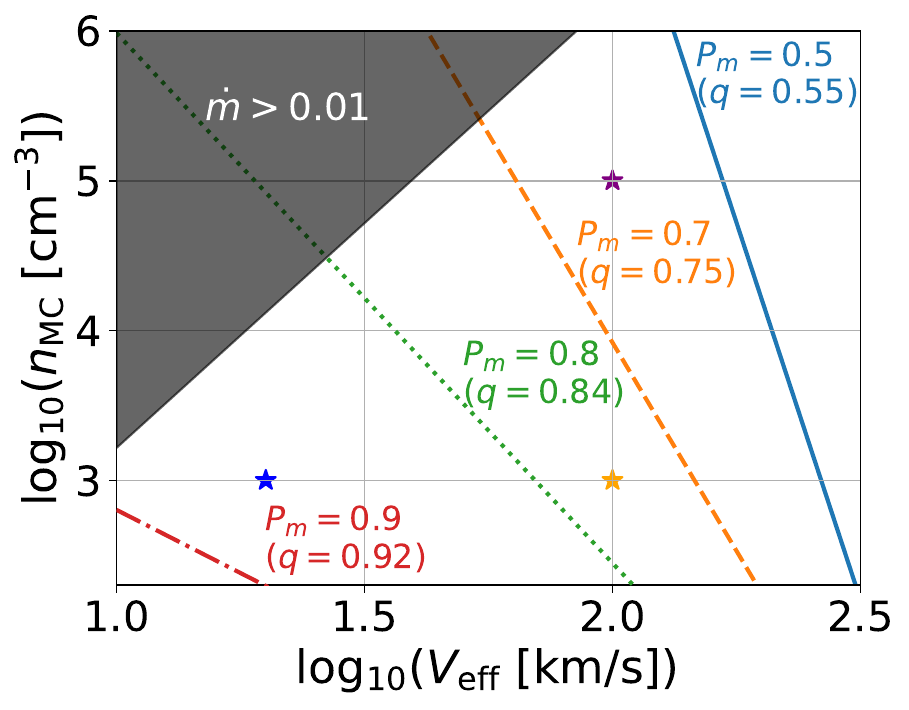}
\caption{
$n_{\mathrm{MC}}$--$V_{\mathrm{eff}}$ parameter space satisfying $\Phi_{\mathrm{H}}\ge\Phi_{\mathrm{MAD}}$ for each magnetic Prandtl number. 
The solid blue, dashed orange, dotted green, and dot-dashed curves represent the contours $\Phi_{\mathrm{H}}=\Phi_{\mathrm{MAD}}$ for $P_m=0.5(q=0.55)$, $P_m=0.7(q=0.75)$, $P_m=0.8(q=0.84)$, $P_m=0.9(q=0.92)$. 
The MAD condition for $P_m<1$ is satisfied toward higher
$n_{\mathrm{MC}}$ and $V_{\mathrm{eff}}$. 
The gray shaded region corresponds to $\dot{m}\equiv \dot{M}c^2/L_{\mathrm{Edd}}>0.01$, where the present model is not applicable. 
The blue, orange, and purple stars show the
$(n_{\mathrm{MC}}, V_{\mathrm{eff}})$
parameter sets adopted for Models A/B, C/D, and E/F, respectively (see Table \ref{tab:model} and Section \ref{sec:radiation}).
\label{fig:mad_criterion}}
\end{figure}

We evaluate the parameter space $n_{\mathrm{MC}}-V_{\mathrm{eff}}$ that satisfies $\Phi_{\mathrm{H}}\ge\Phi_{\mathrm{MAD}}$ for
various values of $P_m$.
We adopt the ranges of $n_{\mathrm{MC}}$ and $V_{\mathrm{eff}}$ motivated by typical molecular clouds $10^2\mathrm{cm^{-3}}\le n_{\mathrm{MC}}\le 10^3\mathrm{cm^{-3}}$
\citep{2010ApJ...723..492R},
dense molecular filaments $10^4\mathrm{cm^{-3}}\le n_{\mathrm{MC}}\le 10^5\mathrm{cm^{-3}}$ 
\citep{2015A&A...584A..91K}, and the typical kick velocity of IBH $10\,\mathrm{km\,s^{-1}}\le V_{\mathrm{eff}}\le 300\,\mathrm{km\,s^{-1}}$\citep{2024ApJ...973....5K,2025PASP..137c4203N}.
Within these parameter ranges, we evaluate whether the MAD condition can be satisfied for IBHs moving in molecular clouds.

Figure \ref{fig:mad_criterion} shows that the MAD condition is satisfied for $P_m\gtrsim1$.
We also see that MAD formation is more difficult for smaller values of $P_m$.
This is because a smaller $P_m$ corresponds to more efficient magnetic diffusion, allowing the magnetic flux to escape more easily.
As a result, IBHs embedded in molecular clouds
generally do not reach the MAD state  for $P_m=0.5$, unless IBHs are very rapidly moving in the cloud.

Our model is not applicable to systems with high accretion rates. This is because we assume aspect ratio $h$ is constant throughout the disk. 
At high accretion rates $\dot{m}=\dot{M}c^2/L_{\mathrm{Edd}}>0.01$, 
the accretion flow is likely composed of an inner RIAF ($h\sim1$) and an outer standard disk($h\ll 1$), based on spectral modeling of black hole X-ray binaries \citep{2019A&A...626A.115M, 2022A&A...657A..11B}. 
We adopt $\dot{m}=0.01$ as the applicability criterion of our model, although lower transition accretion rates of $\dot{m}\gtrsim10^{-3}$ have also been proposed (e.g., \citealt{2021A&A...649A..87G}).

MHD simulation of MRI-driven turbulence shows that $0.5 \le P_m \le 3$ \citep{2009A&A...507...19F}, 
while 3D-GRMHD simulation of MAD suggests $1 \le P_m \le 5$ \citep{2025arXiv251025842J}.
This work shows that MAD formation occurs for $P_m\gtrsim1$. 
Therefore, the range of $P_m$ inferred from GRMHD simulations of MADs falls within the parameter regime where our model predicts MAD formation.
Moreover, the overlap between the MRI-driven turbulence range
($0.5\lesssim P_m\lesssim3$)
and the MAD-forming regime derived here suggests that our MAD criterion may be applicable to MRI-driven accretion disks.

In summary, our results indicate that MAD should be formed for $P_m\gtrsim1$ even if we relax the flux-freezing assumption ($P_m\gg1$) commonly adopted in previous work \citep[e.g.,][]{2017MNRAS.470.3332I,2021ApJ...922L..15K,2025ApJ...981L..36K}. 
However, the MAD formation condition shows a strong dependence on the magnetic Prandtl number: 
as $P_m$ decreases and magnetic diffusion becomes more efficient, the parameter space allowing MAD formation shrinks rapidly.

\section{Photon spectra from IBH Accretion Flows}
\label{sec:radiation}

We first estimate the magnetic field strength relevant for radiation using the results from Section~\ref{sec:mad_criterion}. 
Then, we calculate the emission spectra by both thermal and non-thermal electrons in the accretion flow. We estimate the observed flux, taking into account extinction by molecular clouds.
We then discuss the detectability of isolated black holes based on multiwavelength emission spanning the infrared to X-ray bands.

\subsection{Magnetic field strength in the radiation region}
\label{sec:subsec41}

In this subsection, we describe the calculation of the magnetic field strength related to radiation, $B_{\mathrm{rad}}$, by comparing the magnetic field strength determined by different physical processes. The large-scale magnetic flux discussed in Section \ref{sec:mad_criterion} provides an estimate of the ordered magnetic field, $B_{\mathrm{order}}=\sqrt{B_z^2+B_R^2}$, evaluated by Equations~\eqref{eq:B_R}, \eqref{eq:B_z}, and \eqref{eq:B_R_ver2}. 
If this ordered field is too strong, the accretion dynamics would be governed by the magnetic field, leading to a saturation value of $B_{\rm sat}$ given by Equation (\ref{eq:Bsat}). In contrast, if $B_{\rm order}$ is too weak, the MRI amplifies the turbulent field in the disk, resulting in the MRI-driven value, $B_{\rm MRI}=\sqrt{8\pi N_pk_BT_p/\beta_{\rm MRI}}$, where $\beta_{\rm MRI}$ is the plasma beta determined by the non-linear growth of MRI. MHD simulations suggest $\beta_{\rm MRI}\sim10-100$ \citep[e.g.,][]{2013ASPC..474...59H,2014ApJ...784..121S,2019MNRAS.485..163K,2020ApJ...900..100R}, and we adopt $\beta_{\rm MRI}=100$ in this study.

\begin{table*}
\centering
\caption{Parameter sets used in the radiation calculations and resultant physical quantities.}
\label{tab:model}

\begin{tabular}{ccccccccc}
\hline
\multicolumn{8}{c}{Shared parameters} \\
\hline
$M$ [$M_{\odot}$] &
$r_{\mathrm{in}}$ &
$\lambda_m$ &
$s_w$ &
$s_{\mathrm{inj}}$ &
$\eta_{\mathrm{diff}}$ &
$\beta_{\mathrm{rec}}$ &
$\alpha$ \\
\hline
10 & 3 & 0.10 & 0.5 & 2 & 10 & 0.1 & 0.1\\
\hline
\end{tabular}

\vspace{2mm}

\begin{tabular}{cccccc|cccc}
\hline
\multicolumn{6}{c|}{Model parameters} &
\multicolumn{4}{c}{Resultant physical quantities} \\
\hline
Model &
$P_m$ &
$V_{\mathrm{eff}}$
[$\mathrm{km\,s^{-1}}$] &
$n_{\mathrm{MC}}$
[$\mathrm{cm^{-3}}$] &
$N_H$
[$10^{21}\,\mathrm{cm^{-2}}$] &
$d_L$ [kpc] &
$\dot{m}(R_{\mathrm{in}})$ &
$R_{\mathrm{out}}/R_G$ &
$R_{\mathrm{BHL}}/R_G$ &
$R_{\mathrm{IF}}/R_G$ \\
\hline

A & 0.5 & 20  & $10^3$ & 8.8 & 1 &
$3.7\times10^{-6}$ & $1.2\times10^5$ & $4.5\times10^8$ & $2.0\times10^9$\\
B & 1.0 & 20  & $10^3$ & 8.8 & 1 & $3.7\times10^{-6}$ & $1.2\times10^5$ & $4.5\times10^8$ & $1.5\times10^9$\\
C & 0.5 & 100 & $10^3$ & 8.8 & 1 & $4.3\times10^{-7}$ & $5.9\times10^2$ & $1.8\times10^7$ & $4.8\times10^8$\\
D & 1.0 & 100 & $10^3$ & 8.8 & 1 & $4.3\times10^{-7}$ & $5.9\times10^2$ & $1.8\times10^7$ & $3.7\times10^8$\\
E & 0.5 & 100 & $10^5$ & 14 & 3 & $4.3\times10^{-5}$ & $5.9\times10^2$ & $1.8\times10^7$ & $2.7\times10^8$\\
F & 1.0 & 100 & $10^5$ & 14 & 3 & $4.3\times10^{-5}$ & $5.9\times10^2$ & $1.8\times10^7$ & $2.9\times10^8$\\

\hline
\end{tabular}

\end{table*}

\begin{figure}[t]
\plotone{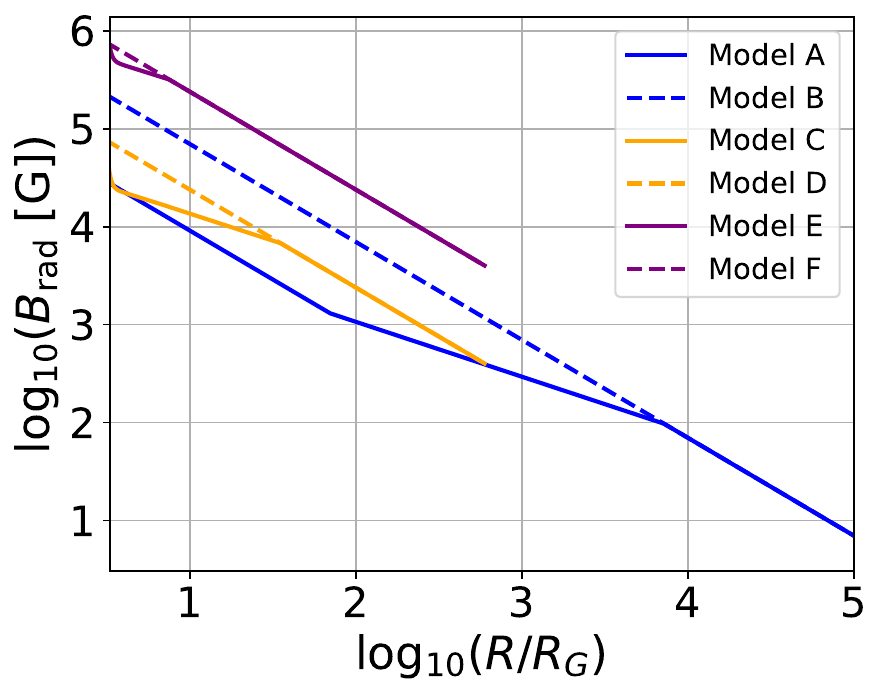}
\caption{The radial profiles of the magnetic field strength used in the radiation calculation, $B_{\mathrm{rad}}$, for Models A--F 
(see Equation~\eqref{eq:B_rad}).
The solid and dashed curves correspond to
$P_m=0.5$ and $P_m=1$, respectively.
\label{fig:B_turb}}
\end{figure}

From the above treatment, 
the magnetic field strength related to radiation is then given by
\begin{equation}
B_{\mathrm{rad}}=
\begin{cases}
B_{\mathrm{MRI}} \, (B_{\mathrm{order}}\le B_{\mathrm{MRI}}), \\
B_{\mathrm{order}} \, (B_{\mathrm{MRI}}\le B_{\mathrm{order}}\le B_{\mathrm{sat}}), \\
B_{\mathrm{sat}} \, (B_{\mathrm{sat}}\le B_{\mathrm{order}}).
\end{cases}
\label{eq:B_rad}
\end{equation}
Figure~\ref{fig:B_turb} shows the radial profile of the magnetic field strength used in the radiation calculation, $B_{\mathrm{rad}}$, 
for Models A--F. 
We consider six representative models 
(A--F; see Table~\ref{tab:model} and Figure \ref{fig:mad_criterion}).
Models A and B are fiducial models, assuming a typical molecular cloud with
$n_{\mathrm{MC}}=10^3\,\mathrm{cm^{-3}}$
and
$V_{\mathrm{eff}}=20\,\mathrm{km\,s^{-1}}$.
Models C and D adopt a higher relative velocity,
$V_{\mathrm{eff}}=100\,\mathrm{km\,s^{-1}}$,
while keeping the same ambient density.
Models E and F adopt the same relative velocity as Models C and D,
$V_{\mathrm{eff}}=100\,\mathrm{km\,s^{-1}}$,
but assume a higher ambient density,
$n_{\mathrm{MC}}=10^5\,\mathrm{cm^{-3}}$,
representative of molecular-cloud filaments and dense cores.
For each set of environmental parameters, we consider both $P_m=0.5$ and $P_m=1$. 
Models A, C, and E adopt $P_m=0.5$, whereas Models B, D, and F adopt $P_m=1$.

In all models, radial distributions of the magnetic fields follow broken-power-law profiles.
For $P_m=0.5$ ($q=0.55<1$; Models A, C, and E), 
the inward transport of magnetic flux is inefficient, leading to a flatter radial profile of the magnetic field and an increase in the plasma beta toward smaller radii. 
This behavior is consistent with the analytical scaling (Equations \eqref{eq:B_z}, \eqref{eq:psi_ana}, and \eqref{eq:B_R_ver2}) $\beta\propto V_{\mathrm{eff}}^{(-11+10q)/3}n_{\mathrm{MC}}^{-13/10}r^{2(q-1)}$.
As a result, the outer region is regulated by the constraint $\beta=\beta_{\mathrm{sat}}$ (Equation (\ref{eq:beta_sat})), while the inner flow becomes less magnetically dominated and approaches a SANE state dominated by MRI turbulence. 
Among these models, Model A exhibits the largest critical radius,
$R_c$, defined by $\beta(R_c)=\beta_{\mathrm{sat}}$.
Compared with Model A, the plasma beta is lower in Models C and E due to their larger $V_{\mathrm{eff}}$. 
In addition, the higher $n_{\mathrm{MC}}$ in Model E further reduces the plasma beta. 
As a result, the critical radius
$R_c$ shifts inward. 
In contrast, for $P_m=1$ ($q=1$; Models B, D, and F), $B_{\mathrm{rad}}$ is constrained by $B_{\mathrm{sat}}$ over the entire radial range.
Since $B_{\mathrm{sat}}\propto V_{\mathrm{eff}}^{-2/3}n_{\mathrm{MC}}^{1/2}$,
Model D has a weaker magnetic field than Model B because of its larger $V_{\mathrm{eff}}$ whereas Model F has a stronger magnetic field than Model B owing to its higher $n_{\mathrm{MC}}$.
The resulting magnetic field profiles
are used to calculate the synchrotron emission from the accretion flow.

\subsection{Thermal Radiation Calculation}
\label{sec:th_rad_calc}

\begin{figure}[t]
\plotone{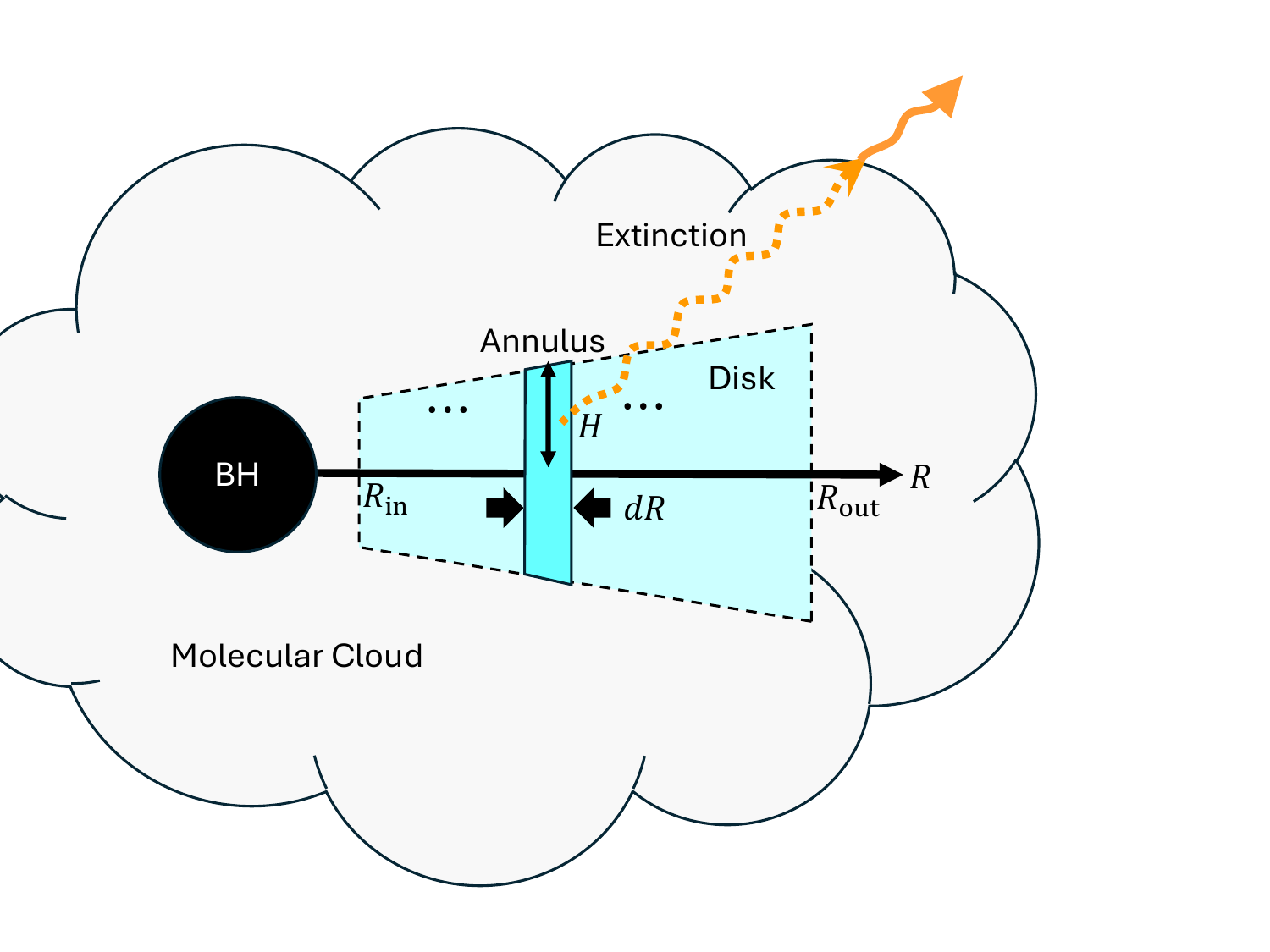}
\caption{Schematic illustration of the radiation calculation from the accretion disk.
The disk is divided into annuli with width dR.
The radiation from each annulus is calculated and integrated over the disk.
The emitted photons suffer extinction in the surrounding molecular cloud.
\label{fig:1Dradiation}}
\end{figure}

We consider thermal radiation from the entire accretion disk.
As shown in Figure~\ref{fig:1Dradiation},
the disk is divided into annuli, and the thermal radiation intensity is calculated for each annulus.
The contributions from all annuli are then summed to obtain the thermal photon spectrum from the entire accretion disk.
For the thermal radiation calculation in each annulus,
we follow the method of \citet{2021ApJ...915...31K}.
Thermal electrons produce multiwavelength emission through thermal synchrotron, bremsstrahlung, and IC.
The thermal radiation strongly depends on the electron temperature.
Therefore, the electron temperature in each annulus must be determined self-consistently.

The heating rate is assumed to originate from viscous dissipation in the accretion flow.
The energy dissipation rate per volume 
is written as
\citep{1994ApJ...428L..13N,1995ApJ...452..710N}
\begin{equation}
\begin{split}
q_{\mathrm{diss}}&=\frac{3\rho \epsilon'|V_R|C_S^2}{2R},\\
\epsilon'&=\frac{\epsilon}{f}=
\frac{1}{f}\frac{5/3-\gamma}{
\gamma-1},
\end{split}
\label{eq:q_diss}
\end{equation}
where
$\rho=m_pN_p$
is the mass density and
$\gamma$
is the adiabatic index.
The parameter
$f$
takes values in the range
$0\le f\le1$;
$f\rightarrow0$
corresponds to a standard thin disk,
whereas
$f\rightarrow1$
corresponds to a RIAF.
To be consistent with
Equations~\eqref{eq:RIAF_V_R}
and \eqref{eq:RIAF_C_S},
we adopt
$\epsilon'\approx1$.
The energy dissipation rate and electron heating rate in each annulus are estimated as
\begin{align}
dQ_{\mathrm{th}}(R)&=q_{\mathrm{diss}}(R)2\pi R\,2H\,dR,\\
dQ_{e,\mathrm{th}}(R)&=f_e(R)
dQ_{\mathrm{th}}(R),
\end{align}
where
$f_e(R)$
is the fraction of dissipated energy transferred to electrons.
To account for electron heating in trans-relativistic magnetic reconnection,
we adopt the electron heating prescription given by \citet{2017ApJ...850...29R,2018MNRAS.478.5209C}
\begin{equation}
f_e(R) \approx \frac{1}{2}\mathrm{exp}\left(-\frac{1-4\beta(R)\sigma_p(R)}{0.8+\sqrt{\sigma_p(R)}}\right),
\label{eq:f_e}
\end{equation}
where $\sigma_p(R)=B_{\mathrm{rad}}(R)^2/4\pi N_p(R)m_pc^2$ is the magnetization parameter.

The cooling rate is obtained by integrating the emitted photon spectrum.
We determine
$T_{e,\mathrm{rad}}$
by iteratively balancing the radiative cooling and electron heating rates.
When radiative cooling is inefficient and cannot balance the heating before the gas accretes onto the IBH,
the electron temperature is instead limited by advection.
We consider two characteristic temperatures,
$T_{e,\mathrm{rad}}$ and
$T_{e,\mathrm{ad}}\approx f_e T_p$ 
(see Equation~\ref{eq:RIAF_T_p}),
and adopt
$T_e=\min\left(T_{e,\mathrm{ad}},T_{e,\mathrm{rad}}\right)$.

\begin{figure}[t]
\plotone{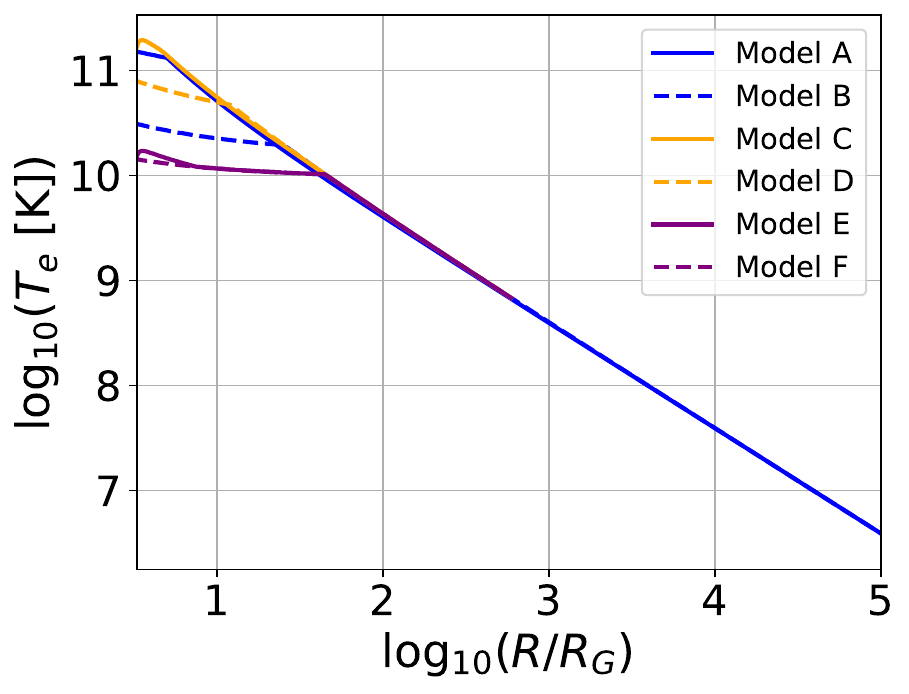}
\caption{The radial profiles of the electron temperature $T_e$ for Models A--F.
The solid and dashed lines correspond to $P_m=0.5$ and $P_m=1$, respectively.
\label{fig:T_e_profile}}
\end{figure}

Figure~\ref{fig:T_e_profile} shows the radial profiles of the electron temperature $T_e$ for Models A--F, obtained by balancing the cooling and heating rates. 
In all models, At large radii, the electron temperature is given by $T_e=T_{e,\mathrm{ad}}$ and follows power-law profile. 
In contrast, at small radii, the electron temperature is determined by radiative cooling ($T_e=T_{e,\mathrm{rad}}$), 
and the radial profile becomes nearly flat. 
Comparing Figures~\ref{fig:B_turb} and \ref{fig:T_e_profile}, models with stronger magnetic fields exhibit lower electron temperatures.
This is because a stronger magnetic field enhances synchrotron cooling, thereby reducing $T_{e,\mathrm{rad}}$.
For $P_m=0.5$ (Models A, C, and E), magnetic flux escapes outward, leading to a weaker magnetic field and less efficient cooling.
For $P_m=1$ (Models B, D, and F), magnetic flux is efficiently transported inward, resulting in a stronger magnetic field and enhanced radiative cooling. 
Therefore, the electron temperature is lower for higher $P_m$.
The transition radius between the advective and radiative regimes increases with the magnetic Prandtl number, reflecting the efficiency of the advective magnetic flux transport. 
The cooling rate also depends on mass accretion rates, and higher mass accretion rates cause more efficient cooling. Thus, the transition radius is larger for Models E and F. 

As shown in Figure \ref{fig:1Dradiation}, 
we also consider the extinction caused by the molecular cloud. 
For optical band, the extinction is dominated by Mie scattering from dust particles in molecular clouds.
We adopt a fitting equation for interstellar extinction across the infrared to ultraviolet wavelengths of the interstellar medium \citep{1989ApJ...345..245C}.
This fitting equation is given with 
$A_V$ and $R_V\equiv A_V/(A_V-A_B)$ 
($A_V$ and $A_B$ are interstellar extinction in V($\lambda=0.55 \mathrm{\mu m}$) and B($\lambda=0.44 \mathrm{\mu m}$) band) as parameter; 
we calculate the extinction assuming $R_V=5$. 
For a given hydrogen column density $N_H$, 
we adopt the relation between $A_V$ and 
the hydrogen column density $N_H$, 
$N_H=2.21\times 10^{21}A_V\,\mathrm{cm^{-2}}$ \citep{2009MNRAS.400.2050G}. 
For soft X-rays, the extinction is dominated by photoelectric absorption by gas and dust in the molecular cloud.
We adopt the fitting formula for the X-ray absorption and scattering cross sections from \citet{2000ApJ...542..914W}.

\subsection{Nonthermal Radiation Calculation}

For the nonthermal component, 
we adopt the one-zone prescription of \citet{2020ApJ...905..178K},
in which the central emission region is treated as a single zone.
We model the nonthermal electron distribution within $R_{\mathrm{NT}}=10R_G$,
motivated by GRMHD simulations 
showing that large-scale magnetic reconnection repeatedly occurs within this radius \citep{2022ApJ...924L..32R}. 
Relativistic magnetic reconnection is known to efficiently generate nonthermal electron distributions,
as demonstrated by recent 3D PIC simulations
\citep{2021ApJ...922..261Z,2023ApJ...956L..36Z}.
We assume that a large fraction of the magnetic energy released by magnetic reconnection is transferred to nonthermal electrons, which is also supported by recent PIC simulations \citep{2023ApJ...946...77H}.
Then, non-thermal particle production power during the reconnection event is estimated to be
$L_{\mathrm{rec}}=\int_{R_G}^{R_{\mathrm{NT}}} U_B \beta_{\mathrm{rec}}cR\Delta\phi dR$,
where $U_B = B_{\mathrm{rad}}^2/8\pi$ is the magnetic energy density, $R\Delta\phi dR$ is the effective area of reconnection regions, and $\beta_{\rm rec}$ is the reconnection efficiency.

Since magnetic reconnection occurs intermittently, the nonthermal emission should be treated using a time-averaged approximation.
The time-averaged nonthermal particle luminosity is given by  
$L_{\mathrm{NT}}=L_{\mathrm{rec}}T_{\mathrm{dur}}/T_{\mathrm{int}}$, where $T_{\mathrm{dur}}$ and $T_{\mathrm{int}}$ are the duration and recurrence interval of reconnection events.
We adopt $\Delta\phi=2\pi/3$, $T_{\mathrm{int}}=2000\,R_G/c=100M_1\,\mathrm{ms}$, and $T_{\mathrm{dur}}=40\,R_G/c=2M_1\,\mathrm{ms}$ based on GRMHD simulations \citep{2022ApJ...924L..32R, 2023MNRAS.526.2924J}. These timescales are shorter than typical integration time of observations, and thus, our time-average treatment is justified.
The reconnection efficiency is set to
$\beta_{\mathrm{rec}}=0.1$,
following PIC simulations
\citep{2014ApJ...783L..21S,2023ApJ...956L..36Z}.
We estimate the magnetic field in the emission region by averaging it with a weight of available magnetic energy:
$\langle B\rangle
=\int_{R_G}^{R_{\mathrm{NT}}} B_{\mathrm{rad}}U_BR\Delta\phi dR/\int_{R_G}^{R_{\mathrm{NT}}}U_BR\Delta\phi dR$.

\begin{figure}[t]
\plotone{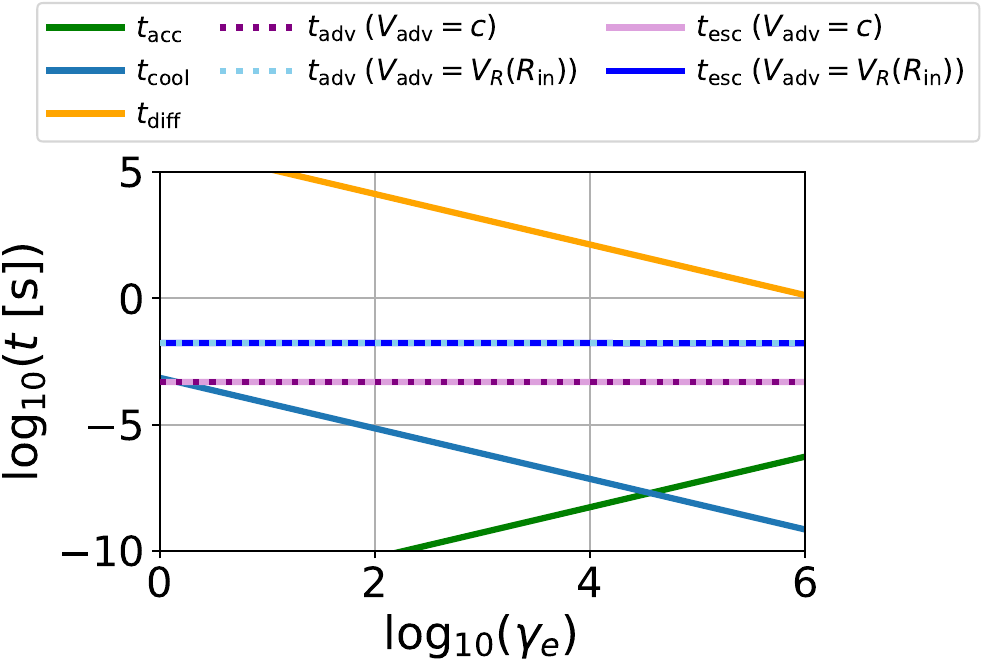}
\caption{Characteristic timescales of nonthermal electrons for Model B.
\label{fig:timescale}}
\end{figure}

\begin{figure}[t]
\plotone{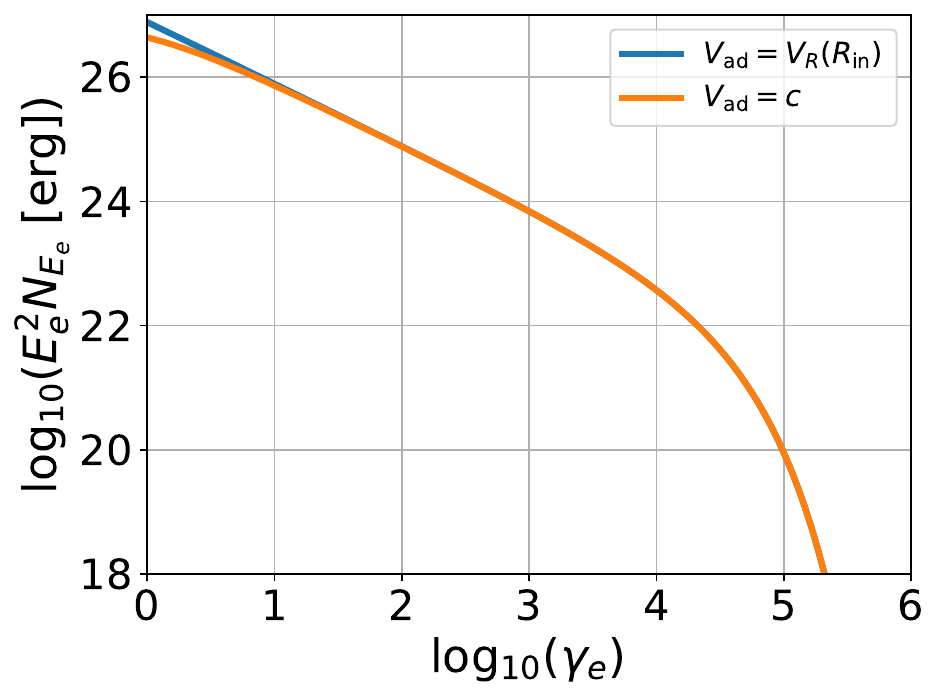}
\caption{Steady-state energy spectrum of nonthermal electrons for Model B.
\label{fig:N_E_e}}
\end{figure}

We solve the transport equation for the non-thermal electrons:
\begin{equation}
\frac{d}{dE_e}\left(\frac{E_eN_{E_e}}{t_{e,\mathrm{cool}}}\right)
-\frac{N_{E_e}}{t_{e,\mathrm{esc}}} + \dot{N}_{E_e,\mathrm{inj}}=0,
\label{eq:trans_NT}
\end{equation}
where $E_e$ is the electron energy, $N_{E_e}$ is the number spectrum, $\dot{N}_{E_e, \mathrm{inj}}$ is the injection term, and $t_{e,\mathrm{cool}}$ and $t_{e,\mathrm{esc}}$
are the cooling and escape timescales, respectively.
We use a single power-law injection function with
an index $s_{\mathrm{inj}}$ and an exponential cutoff at $E_{e,\mathrm{cut}}$, i.e., $\dot{N}_{E_e, \mathrm{inj}}\propto (E_e/E_{e,\mathrm{cut}})^{-s_{\mathrm{inj}}}\,\mathrm{exp}(-E_e/E_{e,\mathrm{cut}})$.
We adopt $s_{\mathrm{inj}}=2$ 
as reference value based on recent 3D PIC simulations \citep{2021ApJ...922..261Z,2023ApJ...956L..36Z}.
The normalization is determined by the nonthermal luminosity $L_{\mathrm{NT}}=\int_{m_ec^2}^{\infty}E_e\dot{N}_{E_e,\mathrm{inj}}dE_e$.
The cutoff energy is obtained by balancing the
acceleration and the shortest loss timescales, $t_{e,\mathrm{acc}}=t_{e,\mathrm{cool}}$.
Figure~\ref{fig:timescale} shows the characteristic acceleration, cooling, and escape timescales of non-thermal electrons for Model B.
The acceleration
timescale is given by $t_{e,\mathrm{acc}}\approx r_{e,L}/c\beta_{\mathrm{rec}}$, where $r_{e,L}=E_e/(e\langle B\rangle)$ is the Larmor radius.
As the cooling process of non-thermal electrons,
we only consider synchrotron emission, $t_{e,\mathrm{cool}}=t_{e,\mathrm{syn}}$,
where $t_{e,\mathrm{syn}}$ is the synchrotron cooling timescale for electrons.
We consider advective (infall toward the IBH) and diffusive escapes. 
The advective and diffusive escape timescales
are given by $t_{\mathrm{adv}}=R_{\mathrm{NT}}/V_{\mathrm{adv}}$ and $t_{e,\mathrm{diff}}=3R_{\mathrm{NT}}^2/(\eta_{\mathrm{diff}}\, r_{e,L}c)$, 
respectively, and 
the escape timescale is given by $t_{e,\mathrm{esc}}^{-1}=t_{\mathrm{adv}}^{-1}+
t_{e,\mathrm{diff}}^{-1}$.
Here, $\eta_{e,\mathrm{diff}}$ characterizes the deviation from the Bohm diffusion limit. 
We adopt $\eta_{e,\mathrm{diff}}=10$.
The advection velocity in the reconnection region, $V_{\mathrm{adv}}$, is uncertain. 
We adopt
$V_{\mathrm{adv}}=c$ as a fiducial value in this work.
Even if the escape process is modified by vertical outflows or a different diffusion prescription, the escape timescale cannot be shorter than the light crossing time, $t_{\mathrm{adv}}=R_{\mathrm{NT}}/c$.
For the adopted parameters, 
$t_{\mathrm{adv}}\ll t_{e,\mathrm{diff}}$, 
and particle escape is therefore dominated by advection.
Figure~\ref{fig:N_E_e} shows the resulting steady-state electron spectrum for Model B.
The spectrum is almost unchanged when adopting
$V_{\mathrm{adv}}=c$
or
$V_{\mathrm{adv}}=V_R(R_{\mathrm{in}})=0.029\,c\,r_{\mathrm{in},0.48}^{-0.5}$,
because
$t_{e,\mathrm{cool}}\ll t_{\mathrm{adv}}$
for the X-ray-emitting electrons.

Similar to the thermal emission, nonthermal X-ray emission is also affected by extinction in molecular clouds.
The extinction of the nonthermal emission is calculated following Section~\ref{sec:th_rad_calc}.

\subsection{Multiwavelength Emission Spectra}

\begin{figure}[t]
\plotone{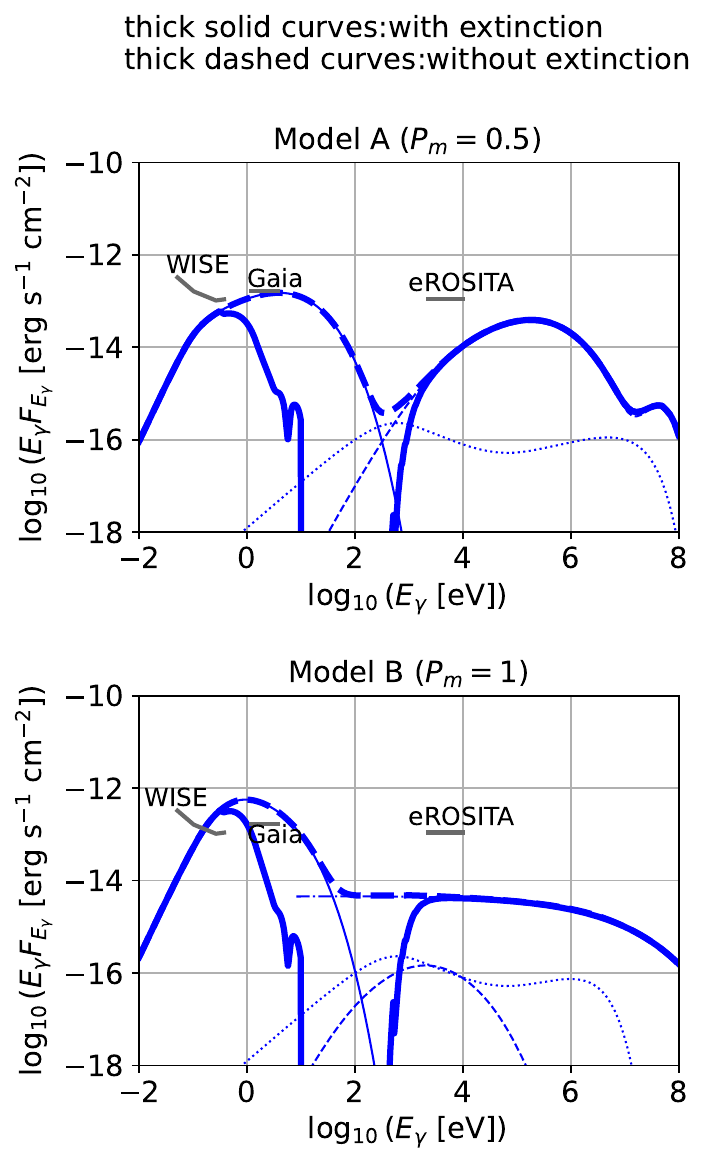}
\caption{
Multiwavelength spectra for Models A (top) and B (bottom).
The thick dashed and solid curves indicate the intrinsic and extincted total spectra, respectively. 
The thin solid, dotted, dashed, and dash-dotted curves represent the intrinsic thermal synchrotron, bremsstrahlung, IC and nonthermal synchrotron components, respectively.
The gray lines are sensitivity limits for
WISE (cryogenic all-sky survey; 5$\sigma$ point-source sensitivities for W1–W4 bands; \citealt{2010AJ....140.1868W}), Gaia (G band; \citealt{2021A&A...649A...1G}), and eROSITA (4 yr; \citealt{2021A&A...647A...1P}).
\label{fig:sed_n1e3_V2e6}}
\end{figure}

We present the multiwavelength emission spectra calculated from the accretion flow model described above, assuming a distance to the IBH of $d_L=1\,\mathrm{kpc}$ (Models A--D).
Figure \ref{fig:sed_n1e3_V2e6} shows multiwavelength spectra from the accretion flow in a typical molecular cloud with $N_H=8.8\times10^{21}\,\mathrm{cm^{-2}}$
\citep{2009ApJ...699.1092H,2017ApJ...834...57M} (Models A and B).
For both models, the thermal synchrotron component is dominant in infrared and optical bands.
Model B ($P_m=1$) satisfies the MAD criterion, whereas Model A ($P_m=0.5$) does not (Figure~\ref{fig:mad_criterion}).
Consequently, the magnetic field strength in Model A is weaker than that in Model B (Figure~\ref{fig:B_turb}), 
resulting in weaker thermal synchrotron emission. 
For Model A, The thermal synchrotron component is not detectable with Gaia and WISE. 
For Model B, the intrinsic thermal synchrotron flux exceeds the Gaia sensitivity limit. However, detection with Gaia is challenging because of strong extinction by the molecular cloud.
In contrast, extinction is much weaker in the infrared band, allowing the source to be detectable with WISE.

For X-ray bands, the dominant component differs between the models. For Model A, IC component is dominant, which appears as a second peak in the spectrum. The strong IC component stems from the higher electron temperature (see Figure \ref{fig:T_e_profile}). The peak energy of the IC component is $\sim 100$ keV, which causes the flux for $\lesssim10$ keV to be lower than the eROSITA sensitivity. In contrast, non-thermal synchrotron component dominates over thermal components for Model B owing to efficient particle acceleration by MAD\footnote{Among the thermal component, bremsstrahlung contributes significantly to the thermal emission in the X-ray to MeV gamma-ray range, which is consistent with the RIAF models for Sgr A* \citep[e.g.,][]{2003ApJ...598..301Y}.}, but the resulting flux is lower than the eROSITA sensitivity.
Nevertheless, sensitive pointing telescopes, such as Chandra and XMM-Newton, might be able to detect these objects for both models. 


\begin{figure}[t]
\plotone{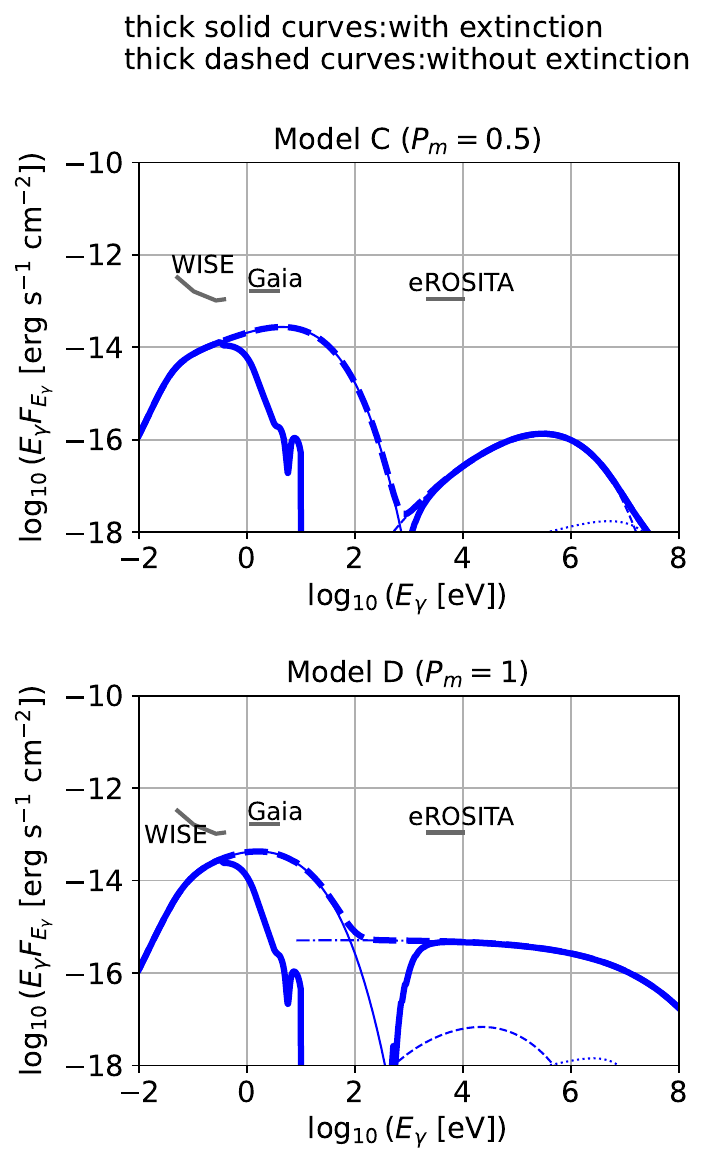}
\caption{Same as Figure~\ref{fig:sed_n1e3_V2e6}, 
but for Models C(top) and D(bottom).
\label{fig:sed_n1e3_V1e7}}
\end{figure}

Figure~\ref{fig:sed_n1e3_V1e7} shows the multiwavelength spectra for Models C and D 
(see Table~\ref{tab:model} for their parameter sets).
Compared with Models A and B shown in Figure~\ref{fig:sed_n1e3_V2e6}, 
the overall emission is suppressed across the entire energy range 
because of 
the strong dependence of the mass accretion rate on the relative velocity, 
$\dot{M}(R\le R_{\mathrm{out}})\propto V_{\mathrm{eff}}^{-4/3}$ (see Equations~\ref{eq:Mdot_0} and \ref{eq:Mdot}).
Therefore, Models C and D
are significantly more difficult to detect. 
The flux ratio of the IC component to the thermal synchrotron component for Model C is much lower than that for Model A;
$E_{\gamma}F_{E_{\gamma},\mathrm{IC,pk}}/E_{\gamma}F_{E_{\gamma},\mathrm{syn,pk}}\simeq0.4$ and $0.01$ for Models A and C, respectively. 
These two models have a similar electron temperature (see Figure \ref{fig:T_e_profile}), whereas the density for Model C is about an order of magnitude lower than that for model A. Thus, the Compton $y$ parameter, $y\propto T_e^2N_p$, is about an order of magnitude lower, leading to a significant difference in the IC component.


\begin{figure}[t]
\plotone{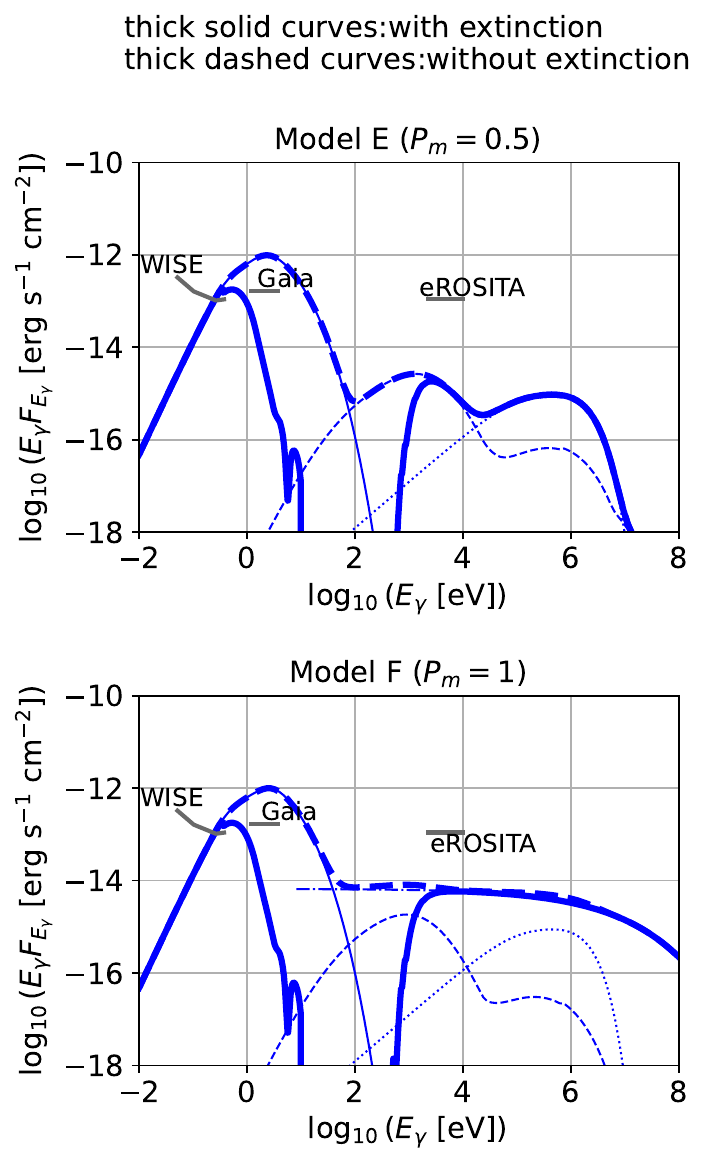}
\caption{Same as Figure~\ref{fig:sed_n1e3_V2e6}, 
but for Models E(top) and F(bottom).
\label{fig:sed_n1e5_V1e7}}
\end{figure}

Figure~\ref{fig:sed_n1e5_V1e7} shows the multiwavelength spectra for Models E and F, which corresponds to IBHs embedded in dense molecular filaments (see Table~\ref{tab:model}). 
Although such regions are significantly denser than typical molecular clouds, 
their smaller characteristic sizes keep the line-of-sight column density comparable. 
Thus, we adopt $N_H=1.4\times10^{22}\,\mathrm{cm^{-2}}$
\citep{2014prpl.conf...27A}.
We adopt a larger distance of $d_L=3\mathrm{kpc}$ for Models E and F because IBHs embedded in dense molecular cloud cores and filaments are expected to be much rarer than those embedded in typical molecular clouds.
Compared to Figure~\ref{fig:sed_n1e3_V1e7}, the emission is enhanced over the entire spectrum.
This enhancement is primarily caused by the higher mass accretion rate,
$\dot{M}(R\le R_{\mathrm{out}})\propto n_{\mathrm{MC}}$
(see Equations~\ref{eq:Mdot_0} and \ref{eq:Mdot}).
However, because of the larger adopted distance ($d_L=3\,\mathrm{kpc}$), the predicted X-ray flux remains below the eROSITA sensitivity for Models E and F.
For Model E, the higher luminosity leads to more efficient radiative cooling and hence a lower electron temperature
(see Figure~\ref{fig:T_e_profile}).
As a result, the IC component becomes substantially weaker and its peak shifts toward lower energies, where extinction becomes more severe.
Moreover, the bremsstrahlung component appears as a distinct third peak at high energies, whereas it is hardly visible in Models A--D, which represent typical molecular cloud environments.
For Model F, the nonthermal synchrotron emission extends from the X-ray to the MeV range ($\sim10^{-14}\,\mathrm{erg\,s^{-1}\,cm^{-2}}$).
If there are IBHs embedded in molecular clouds and cores within $d_L= 1\mathrm{kpc}$, 
their emission could be detectable with eROSITA and future MeV satellites
\citep{2022JATIS...8d4003C,2020APh...114..107A}.


\section{Discussion}
\label{sec:discussion}

\begin{figure}[t]
\plotone{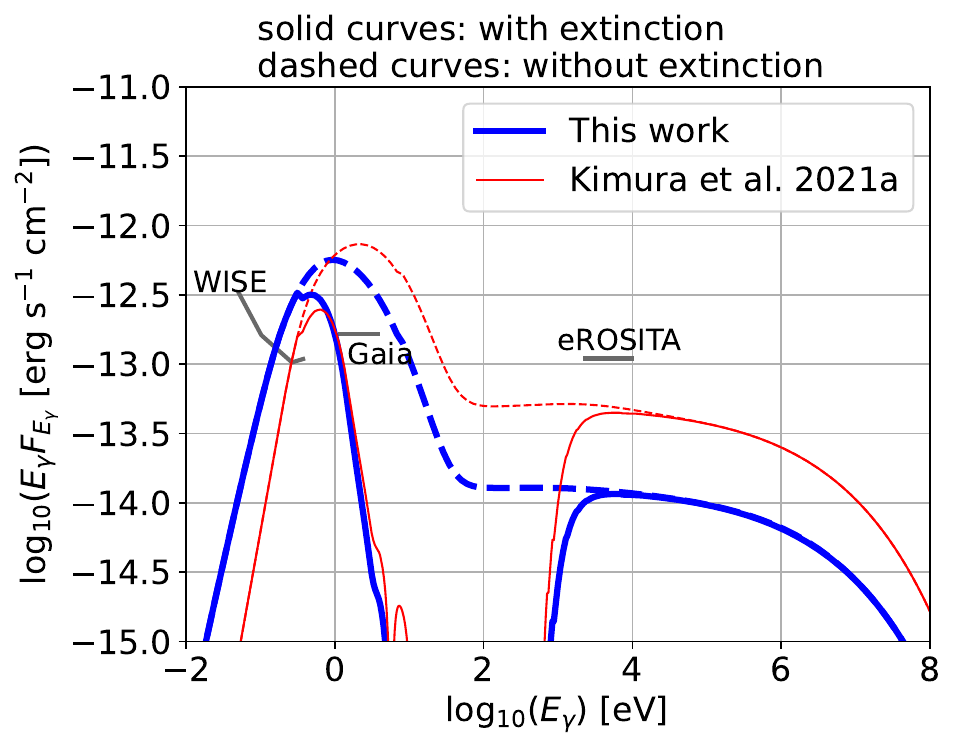}
\caption{Comparison of the multiwavelength spectra for Models B (blue) and the one-zone model of \citealt{2021ApJ...922L..15K} (red).
Solid and dashed curves indicate spectra with and without extinction, respectively.
\label{fig:sed_A_B_SK21}}
\end{figure}

\subsection{Comparison to a previous one-zone model}

We compare our results with those of \citet{2021ApJ...922L..15K}. 
As representative examples, 
we focus on Models B. 
The previous work is a one-zone model at $R=10R_G$ 
for both of the thermal and nonthermal components,
in which the magnetic field $B$ is given by $B=\sqrt{8\pi N_pk_BT_p/\beta}$, 
assuming a constant plasma beta,
$\beta=0.1$,
and a viscous parameter,
$\alpha=0.3$.
In this model, the accretion flow is always in the MAD state. The nonthermal luminosity $L_{\mathrm{NT}}$ and total electron heating rate $Q_{e,\mathrm{th}}$ are given by $L_{\mathrm{NT}}=f_e\epsilon_{\mathrm{NT}}L_{\mathrm{diss}}$ and $Q_{e,\mathrm{th}}=f_e(1-\epsilon_{\mathrm{NT}})L_{\mathrm{diss}}$, respectively, where the dissipated luminosity is defined as $L_{\mathrm{diss}}=\epsilon_{\mathrm{diss}}\lambda_w\dot{M}_{\mathrm{HL}}c^2$.
Here, $f_e$ denotes the electron heating fraction, 
$\epsilon_{\mathrm{NT}}$ is the ratio of nonthermal production to dissipation energy, 
and $\epsilon_{\mathrm{diss}}$ is the ratio of dissipation to accretion energy.
\citet{2021ApJ...922L..15K} adopted $f_e=0.3$, $\epsilon_{\mathrm{diss}}=0.15$, and $\epsilon_{\mathrm{NT}}=0.33$.
While the accretion rate at the emission region is provided using a free parameter, $\lambda_w$, in their model,
we instead 
determine the accretion rate using~\eqref{eq:Mdot} as
$\dot{M}(10R_G)
=\lambda_w\dot{M}_{\mathrm{HL}}
=\lambda_m(10R_G/R_{\mathrm{out}})^{s_w}\dot{M}_{\mathrm{HL}}$.
This procedure assures a fair comparison in the sense that the released energy at the emission region is comparable each other.

Figure~\ref{fig:sed_A_B_SK21} shows a comparison of the multiwavelength spectra for Model B and the one-zone model of \citet{2021ApJ...922L..15K}. 
Compared with the one-zone model, our 1D model produces relatively stronger infrared emission, where extinction by the molecular cloud is much weaker. 
This is because the infrared emission includes contributions from the outer regions of the accretion disk ($R\sim 100R_G$), where the SSA turnover frequency shifts to lower frequencies. 
As a result, the infrared emission in our 1D model is detectable, whereas the one predicted by the one-zone model of \citet{2021ApJ...922L..15K} is unlikely to be detectable with WISE.
In the X-ray band, the model of \citet{2021ApJ...922L..15K} is close to the eROSITA sensitivity limit, suggesting that X-ray detection may be possible under favorable conditions.
However, Model B is less bright than the previous model and 
is therefore unlikely to be detected by eROSITA.
The difference in the nonthermal luminosity 
can be explained by the value of $\epsilon_{\mathrm{NT}}$ and $f_e$.
Using $f_e(10R_G)=0.144$ (see Equation~\ref{eq:f_e}) and 
$\epsilon_{\mathrm{diss}}(10R_G)=0.337$
\footnote{Using Equations~\eqref{eq:RIAF_V_R}, \eqref{eq:RIAF_C_S}, \eqref{eq:RIAF_N_p} and \eqref{eq:q_diss}, 
the dissipation rate per volume can be written as $q_{\mathrm{diss}}\simeq (3/16r)(\dot{M}c^2/\pi R^2 2H)$. 
We obtain $\epsilon_{\mathrm{diss}}(10R_G)=\int_{R_G}^{10R_G}2\pi R2H q_{\mathrm{diss}}dR/\dot{M}c^2=0.337$.}, 
we obtain
$\epsilon_{\mathrm{NT}}
=L_{\mathrm{NT}}/(f_e(10R_G)\epsilon_{\mathrm{diss}}(10R_G)\dot{M}c^2)
=0.072$.
These values are smaller than the value adopted by \citet{2021ApJ...922L..15K}, $\epsilon_{\mathrm{NT}}=0.33$ and $f_e=0.3$, 
naturally leading to a fainter 
nonthermal component in this work.

\subsection{Feedback by ionizing photons}

High-energy photons emitted from the accretion flow around the IBH may ionize the weakly ionized plasma upstream of the bow shock (see Figure~\ref{fig:accretion_disk}).
If this occurs, the mass accretion rate may be suppressed \citep{2011ApJ...739....2P, 2012ApJ...747....9P, 2013ApJ...767..163P, 2020MNRAS.495.2966S, 2024MNRAS.528.2588O}. 
To assess the impact of radiative feedback, 
we estimate the ionization-front radius,
$R_{\mathrm{IF}}$,
which corresponds to the boundary between the photoionized region and the neutral ambient medium.
The accretion flow model described above adopts the Hoyle–Lyttleton radius,
$R_{\rm HL}=2GM/V_{\rm eff}^2$.
Radiative feedback becomes important when
$R_{\mathrm{IF}} > R_{\mathrm{HL}}$. 
By adopting the analytical formulation of \citet{2020MNRAS.495.2966S}, 
$R_{\mathrm{IF}}$ is given by
\begin{equation}
\begin{split}
&R_{\mathrm{IF}} = 
\mathrm{min}\left(r_{\mathrm{Strm}},\, r_{\mathrm{neutral-flow}}\right), \\
& r_{\mathrm{Strm}} = \left(\frac{3\dot{N}_{\mathrm{ion}}}{4\pi n_{\mathrm{MC}}^2\alpha_{\mathrm{B}}}\right)^{\frac{1}{3}}, \\
& r_{\mathrm{neutral-flow}} = \left(\frac{\dot{N}_{\mathrm{ion}}}{4\pi n_{\mathrm{MC}}V_{\mathrm{eff}}}\right)^{\frac{1}{2}},
\label{eq:R_IF}
\end{split}
\end{equation}
where $r_{\mathrm{Strm}}$ is the Strömgren radius determined by the balance between photoionization and recombination, 
$r_{\mathrm{neutral-flow}}$ is the ionization-front radius obtained by balancing the ionizing photon production rate with the flux of gas supplied to the accretion flow across the ionization front, 
$\dot{N}_{\mathrm{ion}}$ is the ionizing photon emission rate, and $\alpha_{\mathrm{B}}\approx 3.3\times 10^{-13}\,\mathrm{cm^{3}\,s^{-1}}$ is the Case-B recombination coefficient for
$T_0=10^4\,\mathrm{K}$ \citep{1992ApJ...387...95F}, where $T_0$ is the characteristic temperature of the gas photoionized by radiation from the accretion flow at the upstream of the bow shock.
The ionizing photon emission rate is estimated as 
\begin{equation}
\dot{N}_{\mathrm{ion}}=\int_{E_{\gamma,0}}^{\infty} \frac{E_{\gamma}F_{E_{\gamma}}4\pi d_L^2}{E_{\gamma}^2} dE_{\gamma},
\end{equation}
where $E_{\gamma,0}=13.6\,\mathrm{eV}$ is the ionization threshold energy, and $E_{\gamma}F_{E_{\gamma}}$ is the radiation flux from the accretion flow without molecular cloud extinction (see Figures~\ref{fig:sed_n1e3_V2e6}--\ref{fig:sed_n1e5_V1e7}). 
The size of the molecular cloud, $R_{\mathrm{MC}}$, is $R_{\mathrm{MC}}\sim10\,\mathrm{pc}\gg R_{\mathrm{HL}}$, 
and the effect of extinction is therefore assumed small.
We evaluate Equation~\eqref{eq:R_IF} for Models A-F and find that
$R_{\mathrm{IF}} > R_{\mathrm{HL}}$
in all cases (see Table~\ref{tab:model}), 
suggesting that radiative feedback may affect the mass accretion rate.
However, the accretion rate is already reduced by outflows in this work, decreasing the luminosity of the central source and thus weakening the radiative feedback.
As a result, the feedback is expected to be self-regulated, and the net suppression of the mass accretion rate may be modest.

\subsection{Prospects for identifying IBHs}

In Section~\ref{sec:radiation}, we extended the one-zone model to a 1D model that includes magnetic flux transport.
This enabled us to more accurately estimate the infrared emission, which is underestimated in the one-zone model \citep{2021ApJ...922L..15K}, allowing us to assess the detectability of IBHs through multi-wavelength infrared and X-ray observations.
Over a wide range of parameters,  IBHs can be detected using infrared emission. 
However, IBHs in typical molecular clouds are unlikely to be detected in X-rays even with eROSITA, one of the most sensitive all-sky X-ray surveys.
Nevertheless, the predicted X-ray flux is detectable with deeper pointing telescopes such as Chandra and XMM-Newton. Thus, X-ray emitting IBHs could exist in archival data of these telescopes targeting 
nearby molecular clouds, such as Taurus and Orion.

Even when infrared and X-ray emission from IBHs are detectable,
their identification remains challenging because other sources, such as protostars, white dwarf binaries, and AGNs, can also produce emission in these bands.
Therefore, additional observational signatures are required to distinguish IBHs from other sources.
One promising method is short-timescale variability originating near the black hole.
In the vicinity of a stellar-mass black hole, electromagnetic emission can exhibit variability on sub-second timescales, reflecting the short dynamical timescale
(e.g., accretion and orbital timescales) 
near the black hole.
In contrast, protostars, white dwarf binaries, and AGNs are generally expected to vary on longer characteristic timescales of 10--1000 sec.
This difference in variability timescales may provide a key observational signature for identifying IBHs. 
A promising strategy is as follows: First, we pre-select IBH candidates in the Milky Way using multiwavelength SED properties predicted in our 1D emission model. Then, we can perform dedicated follow-up observations in X-ray and infrared to search for sub-second scale variability.

Another feature that IBHs are distinguishable is MeV gamma-ray emission. Nonthermal synchrotron emission from MADs surrounding IBHs extends from X-ray to MeV gamma-rays, and the flux of these two bands are comparable. These MeV gamma-rays from IBHs in dense molecular filaments could be detectable with future MeV satellites \citep[e.g.,][]{2020APh...114..107A,2022JATIS...8d4003C}. Since other astrophysical objects, such as accreting WDs, protostars, and AGN, are not expected to emit MeV gamma-rays as luminous as X-ray bands. Therefore, we can identify IBHs if we discover bright MeV gamma-rays from nearby molecular clouds whose flux is comparable to the X-ray source inside that molecular clouds.

In addition to the multi-wavelength spectrum discussed above, the absence of orbital variability may also help distinguish isolated black holes from accreting binary systems. Persistent emission associated with accretion and outflows would therefore provide an additional observational signature of IBHs (e.g., \citealt{2025A&A...700A..49M}).

\subsection{IBHs as unidentified gamma-ray sources}


Previous studies suggest that IBHs embedded in molecular clouds may contribute to PeV cosmic rays and be the origins of unidentified gamma-ray sources in GeV--PeV bands \citep{2012MNRAS.427..589B, 2017MNRAS.470.3332I, 2025ApJ...985..251K, 2025ApJ...981L..36K}. In such a scenario, MADs, associated BH magnetospheres, and jets  are accelerating leptons and hadrons up to GeV--PeV energies, and these nonthermal particles produce gamma rays. The luminosity of these gamma rays should be comparable or less than the infrared and X-ray luminosities from the MADs based on energetics and radiation efficiency. Then, based on our multiwavelength spectrum calculations, we should detect X-ray and infrared-emitting objects within the error regions of unidentified GeV--PeV gamma-ray sources. Multi-wavelength analyses are essential to determine whether IBHs in molecular clouds are the origins of unidentified GeV--PeV gamma-ray sources.

\section{Summary} 
\label{sec:summary}

In this paper, we investigated the detectability of IBHs through multiwavelength emission from accretion flows embedded in molecular clouds. As an IBH moves through a molecular cloud, a bow shock forms around the BH and the shocked gas is captured to form an accretion flow. 
We assumed that magnetic flux is transported from the bow shock to the outer edge of the accretion flow through flux freezing.
Within the accretion flow, we numerically solved the 1D magnetic flux transport equation, including both advection and diffusion of large-scale magnetic flux. 
The magnetic flux was constrained between two physically motivated limits: a MAD limit at high magnetization, where magnetic pressure plays a crucial role in dynamics, and an MRI-turbulence limit at low magnetization, corresponding to a maximum plasma beta of $\beta_{\rm MRI}=100$. Based on the resulting magnetic flux, we evaluated the conditions for MAD formation and calculated the resulting multiwavelength emission spectra.
We found that MAD states are generally achieved for $P_m\gtrsim1$. For $P_m<1$, 
MAD formation is restricted to a limited parameter range (Figure~\ref{fig:mad_criterion}). 
Since our nonthermal emission model assumes particle acceleration by magnetic reconnection in MAD accretion flows, the MAD criterion effectively determines the parameter range in which detectable X-ray and MeV gamma-ray emission can be expected.

Compared to previous one-zone models, 
our 1D model predicts enhanced infrared emission by thermal synchrotron radiation due to contributions from the outer regions of the accretion flow ($R\sim100R_G$) improving the detectability with WISE. Optical emission is strongly suppressed by molecular cloud extinction.
The X-ray properties depend sensitively on the magnetic Prandtl number $P_m$.
For $P_m=1$ (MAD), the X-ray emission is dominated by synchrotron radiation from nonthermal electrons accelerated by magnetic reconnection, whereas for $P_m=0.5$ (SANE), inverse Compton emission is dominant.
Although eROSITA detection of IBHs in typical molecular clouds remains challenging, the detection is possible if IBHs are embedded in dense molecular cloud filaments and cores within $d_L= 1\mathrm{kpc}$, especially when the accretion flow reaches the MAD state, i.e., $P_m\gtrsim1$ is realized.

These results demonstrate that magnetic-flux transport is an essential ingredient in predicting the multiwavelength observational signatures of IBHs.
The 1D radiation model developed in this work provides a framework for selecting promising IBH candidates from infrared and X-ray surveys.
Future searches for IBHs may benefit from coordinated multiwavelength observations combined with the detection of variability on sub-second timescales, which may provide an additional signature of accretion onto IBHs.

\begin{acknowledgments}
We thank Hirofumi Noda , Kengo Tomida, and Kin Koki for helpful discussions and valuable comments.
S.S.K. acknowledges support by KAKENHI Nos. 23H04899, 26K00733, 26K00696, and the Tohoku Initiative for Fostering Global Researchers for Interdisciplinary Sciences (TI-FRIS) of MEXT’s Strategic Professional Development Program for Young Researchers.
\end{acknowledgments}

\bibliography{references}
\bibliographystyle{aasjournalv7}



\end{document}